\documentclass[lettersize,journal]{IEEEtran}

\usepackage{algorithmic}
\usepackage{algorithm}
\usepackage{array}
\usepackage[caption=false,font=normalsize,labelfont=sf,textfont=sf]{subfig}
\usepackage{textcomp}
\usepackage{stfloats}
\usepackage{url}
\usepackage{verbatim}
\usepackage{graphicx}
\usepackage{cite}
\usepackage{bm}
\usepackage{tikz}
\usetikzlibrary{arrows.meta,calc,decorations.pathreplacing}
\usepackage{xcolor}
\usepackage{amsthm}
\usepackage{amsmath}
\usepackage{balance}

\input{macros/macros.sty}

\usepackage{cleveref}

\newcommand{\dataset}[1]{\texttt{#1}}
\usepackage{xspace}
\newcommand{\tsvdmi}{t-SVDM-I\xspace}
\newcommand{\tsvdmii}{t-SVDM-II\xspace}
\newcommand{\pystarm}{PyStarM\xspace}
\newcommand{\tuckermpi}{TuckerMPI\xspace}
\newcommand{\distarm}{StarMPI-SVD\xspace}



\newcommand{\setcrefnames}[5]{%
  \crefname{#1}{#2}{#3}%
  \Crefname{#1}{#4}{#5}%
}

\setcrefnames{algorithm}{Alg.}{Algs.}{Algorithm}{Algorithms}
\setcrefnames{figure}{Fig.}{Figs.}{Figure}{Figures}
\setcrefnames{table}{Tab.}{Tabs.}{Table}{Tables}
\setcrefnames{section}{Sec.}{Secs.}{Section}{Sections}

\newcommand{\ignore}[1]{}

\begin{document}

\title{StarMPI-SVD: Distributed-Memory Parallel Star-M Tensor SVD} % Grey's proposed title based on Taufique's suggestion, up for discussion
%\title{Fast parallel algorithms for large-scale tensor compression using star-M singular value decomposition on distributed architectures.}

\author{%
Naman~Pesricha\IEEEauthorrefmark{1},
Grey~Ballard\IEEEauthorrefmark{2},
Srinivas~Eswar\IEEEauthorrefmark{3},
Md~Taufique~Hussain\IEEEauthorrefmark{2},
Vishwas~Rao\IEEEauthorrefmark{3},
and Phani~Motamarri\IEEEauthorrefmark{1}%
\thanks{\IEEEauthorrefmark{1}Indian Institute of Science, Bengaluru, KA, INDIA}%
\thanks{\IEEEauthorrefmark{2}Wake Forest University, Winston-Salem, NC, USA}%
\thanks{\IEEEauthorrefmark{3}Argonne National Laboratory, Lemont, IL, USA}%
% \thanks{Corresponding author: Phani~Motamarri (e-mail: [phanim@iisc.ac.in].}%
}

% \author{Anonymous}
        % <-this % stops a space
%\thanks{This paper was produced by the IEEE Publication Technology Group. They are in Piscataway, NJ.}% <-this % stops a space
%\thanks{Manuscript received April 19, 2021; revised August 16, 2021.}}



\maketitle

\begin{abstract}
The star-M tensor algebra generalizes the optimal compression guarantees of the matrix singular value decomposition (SVD) to higher-order tensors.
The \tsvdmi decomposition is computed using a set of invertible transformations and slice-wise truncated SVDs with a prescribed rank, and its adaptive, multirank variant \tsvdmii can determine slice-wise ranks to satisfy a prescribed error tolerance.
Existing implementations of these algorithms are limited to shared-memory environments, restricting their applicability to modern large-scale datasets.
In this work, we present the first distributed-memory implementations of \tsvdmi and \tsvdmii, which are generalized to $N$-way tensors.
To address computational and communication bottlenecks, we design asynchronous, non-blocking communication strategies for efficient data redistribution to obtain highly efficient key computational kernels.
Strong-scaling experiments demonstrate the efficiency and scalability of our approach (at least 70\% parallel efficiency on up to 128 nodes, or 6144 cores), and we apply the star-M compression techniques to a set of application datasets of size up to 768 GB.
\end{abstract}


% !TEX root = ../main.tex
\section{Introduction}

\IEEEPARstart{L}{arge} scientific datasets are often naturally multidimensional.
A flow simulation, for example, stores several velocity components over three spatial dimensions and time.
Flattening such data into a matrix loses this structure, so tensor decompositions are a natural choice for analyzing or compressing it.
The most widely used are CP \cite{hitchcock1927expression,harshman1970foundations,carroll1970analysis}, Tucker \cite{tucker1966some,de2000multilinear}, and Tensor-Train (TT) \cite{Oseledets2011TensorTrainD}.
None of them, however, has the property that makes the matrix SVD so appealing.
By the Eckart--Young theorem, truncating the matrix SVD gives the best possible low-rank approximation \cite{Eckart1936TheAO,mirsky1960symmetric}, and no comparable result holds for these tensor decompositions \cite{Newman2025VariableProjection,delathauwer2000bestrank}.

As we describe in \cref{sec:background}, the $\star_M$ framework closes this gap \cite{kernfeld2015tensor,kilmer2021tensor}.
The idea is to transform the tensor along all but its first two modes and then treat each frontal slice as an ordinary matrix.
With a suitable transform, such as the discrete cosine transform, truncating the SVDs of these slices gives the best approximation of a given t-rank or multi-rank in the Frobenius norm \cite{kilmer2021tensor}.
There are two variants \cite{newman2019step,kilmer2021tensor}: \tsvdmi keeps the same rank in every slice, while \tsvdmii chooses the rank of each slice to meet an error tolerance.

Software has not kept up with this theory.
CP \cite{battaglino2018practical,ma2021efficient,phipps2019software,smith2016medium,EH+21,WKP25}, Tucker \cite{ballard2020tuckermpi,austin2016parallel,KU16}, and TT \cite{daas2022parallel,RTB22} all have scalable parallel libraries, while most $\star_M$ codes are written in MATLAB or Python \cite{Lu18,Newman2025VariableProjection,KVX21}.
Recently, a library called \pystarm was released that includes high performance and parallel implementations of these algorithms \cite{hussain2026highperformancestarmsvdbig}, but it is limited to shared-memory parallelism within a single node.

This paper presents \distarm, the first distributed-memory implementation of \tsvdmi and \tsvdmii, for tensors with any number of modes.
The $\star_M$ framework yields computational steps like tensor-times-matrix (TTM) products and matrix SVDs that are inherently parallelizable, but the difficulty lies in the steps around them, which we address in three ways.
First, we partition the tensor so that the TTMs need no communication (\Cref{sec:partitioning}).
Second, performing the SVDs with no communication requires a different data layout from the TTMs, forcing an all-to-all exchange of the whole tensor.
We break this exchange into non-blocking batches and overlap SVD computation while the data is moving (\Cref{sec:alg:overlap}), which, for example, makes the full run up to $1.55\times$ faster on a climate dataset at 32 nodes.
Third, \tsvdmii needs a global cutoff across all the singular values.
Instead of gathering them on one process, we use a two-level histogram that gathers only a small fraction of them (\Cref{sec:truncation}), which is more than 100 times faster.

In \cref{sec:results}, we test \distarm on five scientific datasets of up to 768\,GB and observe effective strong scaling: every dataset keeps at least 70\% parallel efficiency at the largest node count (up to 128 nodes, or 6144 cores).
On a single node, \distarm runs about as fast as \pystarm.
It is also comparable to the distributed-memory parallel library \tuckermpi \cite{ballard2020tuckermpi} in both compression and runtime at all node counts, and unlike \tuckermpi it has no processor grid to tune.
Thus, \distarm removes the barrier of data set size and enables the extension of the $\star_M$ framework to processing very large data, including those produced by supercomputer simulations.

%\Cref{sec:background} gives the background, \Cref{sec:methodology-new} describes our parallel algorithms, \Cref{sec:results} presents our results, and \Cref{sec:conclusions} concludes the paper.

% !TEX root = ../main.tex

\section{Background}
\label{sec:background}

%\MTH{Srini, can you please own this and the \cref{sec:dist-tsvdm-algo}?}

%Following the \texttt{pystarm} paper for this.
%\begin{enumerate}
%	\item Notation (move TTM, tensor ordering here).
%	\item $\starM$ product and framework.
%	\item t-SVD and its variants. (note on optimality)
%	\item Other relevant background work (software and other tensor libraries)
%\end{enumerate}

\subsection{Notation}
\label{sec:notation}
We denote scalars by lowercase italic letters (e.g., $x$), vectors by boldface lowercase (e.g., $\V{x}$), matrices by boldface uppercase (e.g., $\mathbf{X}$), and higher-order tensors ($N \ge 3$) by bold calligraphic letters (e.g., $\T{X}$). An $N$-way tensor $\T{X} \in \Rn{d_1 \times \cdots \times d_N}$ has modes of size $d_k$ accessed via indices $i_k$. Matlab-style indexing is used to access these objects. The Frobenius norm, defined as $\norm{\T{X}}_F = \sqrt{\sum x_{i_1 \dots i_N}^2}$ \cite{kolda2009tensor}, serves as the default measure of magnitude throughout this work. The mode-$k$ unfolding $\Mz{X}{k}$ arranges the mode-$k$ fibers of $\T{X}$ as columns of a matrix. The mode-$k$ tensor-times-matrix product (TTM), written as $\T{Y} = \T{X} \times_k \M{A}$ for $\M{A} \in \Rn{J \times d_k}$, is defined as $\Mz{Y}{k} = \M{A}\Mz{X}{k}$. The resulting tensor $\T{Y}$ has the same dimensions as $\T{X}$ except for mode-$k$ which has size $J$.

Tensors are assumed to be stored in natural-order layout, which is a generalization of the column-major order \cite{Ballard_Kolda_2025}.
We reserve the index $f_i$ for frontal slices, which are obtained by fixing all but the first two indices of a tensor. The trailing indices are linearized via some function $\mathbb{L}(\cdot)$, for e.g., $\M{X}_{f_i} = \T{X}^{f_i} = \T{X}(:, :, i_3, \dots, i_N) \in \Rn{d_1 \times d_2}$ where $f_i = \mathbb{L}(i_3, \dots, i_N)$. The set of all frontal slices is denoted as $F$ whose size is $\abs{F} = \prod_{i=3}^n d_i$. For distributed objects, locally owned pieces are distinguished via a bar subscript like $\Tl{X}$ for the global $\T{X}$.

\subsection{The $\starM$-Product Framework}
\label{subsec:starm-framework}
The $\starM$-framework is a family of matrix-mimetic tensor algebras developed to process data in its inherent multidimensional format \cite{kilmer2021tensor}. Core to this framework is the $\starM$-product , between tensors of appropriate dimensions, parameterized by a set of transformation matrices $\Curly{\M{M}_i \in \Cn{d_i \times d_i}}_{i=3}^N$. 
Given $\T{A} \in \Rn{m \times n \times d_3 \times \cdots \times d_N}$ and $\T{B} \in \Rn{n \times p \times d_3 \times \cdots \times d_N}$ we compute 
$\T{C} = \T{A} \starM \T{B} \in \Rn{m \times p \times d_3 \times \cdots \times d_N}$ in three stages. 
First, we transform both $\T{A}$ and $\T{B}$ via TTMs to the $\M{M}$-domain, via $\T[\widehat]{A} = \T{A} \times_3 \M{M}_3 \cdots \times_N \M{M}_N$ (and similarly for $\T[\widehat]{B}$). 
Then the frontal slices are multiplied independently, denoted by $\triangle$, in the tranform domain $\T[\widehat]{C} = \T[\widehat]{A}\triangle\T[\widehat]{B}$. 
Finally, the result is mapped back to the original domain via $\T{C} = \T[\widehat]{C}\times_3 \M{M}_3^{-1} \cdots \times_N \M{M}_N^{-1}$. 

Under this algebra, the $\starM$-identity tensor $\T{I}$ and the $\starM$-transpose $\T{A}^\Tra$ are defined such that their frontal slices in the transform domain correspond to the standard matrix identity $\M{I}_m$ and matrix transpose $\M[\widehat]{A}_f^\Tra$, respectively.  Since $\starM$-product operations decouple into standard matrix multiplications across frontal slices in the transform domain, classical linear algebra concepts such as orthogonality, inverses, and the singular value decomposition (SVD) extend seamlessly to higher-order tensors. This allows matrix algorithms to be rigorously lifted to the tensor space, forming the foundation for the \tsvdmi and \tsvdmii decompositions.

\subsection{Tensor SVD in the $\starM$ Framework}
We provide a brief overview of the two tensor singular value decompositions (t-SVD) considered in this work. We refer the reader to the following works for more details \cite{kilmer2011factorization,newman2019step,kilmer2021tensor}.

\paragraph{t-SVDM-I}
Given a third-order tensor
$\T{A} \in \Rn{d_1 \times d_2 \times d_3}$ and an invertible $\M{M} \in \Rn{d_3 \times d_3}$,
the full t-SVDM---which serves as the tensor analogue of the full matrix SVD under the $\starM$ framework---is defined as
\begin{equation}
    \T{A}=\T{U}\starM\T{S}\starM\T{V}^\Tra,
\end{equation}
where
$\T{U} \in \Rn{d_1 \times d_1 \times d_3}$
and
$\T{V} \in \Rn{d_2 \times d_2 \times d_3}$
are orthogonal tensors under the $\starM$-product, and
$\T{S} \in \Rn{d_1 \times d_2 \times d_3}$
is an f-diagonal tensor, meaning that each frontal slice
$\T{S}(:,:,i)$
is diagonal for
$i = 1,\dots,d_3$.

The t-SVDM is computed by first applying the transform $\M{M}$ along the third mode, performing independent matrix SVDs on the transformed frontal slices, and then applying the inverse transform to recover the factor tensors in the original domain.
This formulation naturally extends to higher-order tensors
$\T{A} \in \Rn{d_1 \times d_2 \times \cdots \times d_N}$
by applying the transform along modes $3$ through $N$, after which independent matrix SVDs are computed on the resulting transformed frontal slices.

\begin{algorithm}[hbtp]
	\caption{$N$-Way t-SVDM-I Decomposition}\label{alg:tsvdmi}
	\begin{algorithmic}
		\REQUIRE Input tensor $\T{A} \in \Rn{d_1 \times \cdots \times d_N}$, transformation matrices $\Curly{\M{M}_i \in \Rn{d_i \times d_i}}_{i=3}^N$, and truncation rank $k$.
		\ENSURE $\pr{\T{U}_k, \T{S}_k, \T{V}_k}$ with t-rank $k$ such that $\T{A} \approx \T{U}_k \starM \T{S}_k \starM \T{V}_k^\Tra$.
		\STATE {\textsc{t-SVDM-I}}$(\T{A}, \Curly{\M{M}_i}, k)$
		\STATE $\T[\widehat]{A} \gets \T{A} \times_3 \mathbf{M}_3 \times_4 \cdots \times_N \mathbf{M}_N $
		\FOR{$f_i \in F$}
		\STATE $\M[\widehat]{U}, \M[\widehat]{S}, \M[\widehat]{V} \gets \textsc{SVD}\pr{\M[\widehat]{A}_{f_i}}$
		%\STATE $\T[\widehat]{U}^{f_i}, \T[\widehat]{S}^{f_i}, \T[\widehat]{V}^{f_i} \gets \M{U}(:,1:k), \M{S}(1:k,1:k), \M{V}(:,1:k)$
		\STATE $\T[\widehat]{U}_{k}^{f_i}, \T[\widehat]{S}_k^{f_i}, \T[\widehat]{V}_k^{f_i} \gets \textsc{truncate}\pr{\M[\widehat]{U}, \M[\widehat]{S}, \M[\widehat]{V}, k}$
		\ENDFOR
		\STATE $\T{U}_k \gets \T[\widehat]{U} \times_3 \mathbf{M}_3^{-1} \times_4 \cdots \times_N \mathbf{M}_N^{-1}$
		\STATE $\T{S}_k \gets \T[\widehat]{S} \times_3 \mathbf{M}_3^{-1} \times_4 \cdots \times_N \mathbf{M}_N^{-1}$
		\STATE $\T{V}_k \gets \T[\widehat]{V} \times_3 \mathbf{M}_3^{-1} \times_4 \cdots \times_N \mathbf{M}_N^{-1}$
		\RETURN $\pr{\T{U}_k, \T{S}_k, \T{V}_k}$
	\end{algorithmic}
\end{algorithm}

The higher order analogue of the truncated matrix SVD is the truncated t-SVDM-I (\cref{alg:tsvdmi}). By retaining only the top-$k$ lateral slices corresponding to the $k$ largest singular values in the transform domain—where $k < \min(d_1, d_2)$ is the target rank—we obtain the approximation:

\begin{equation}
    \T{A}_k = \T{U}_k \starM \T{S}_k \starM \T{V}_k^\Tra,
\end{equation}
where the reduced factors have dimensions $\T{U}_k \in \Rn{d_1 \times k \times d_3 \times \cdots \times d_N}$, $\T{S}_k \in \Rn{k \times k \times d_3 \times \cdots \times d_N}$, and $\T{V}_k \in \Rn{d_2 \times k \times d_3 \times \cdots \times d_N}$. $k$ is referred to as the \emph{t-rank} of $\T{A}_k$ which is the number of nonzero tubes in the core-tensor $\T{S}$.

\begin{figure*}[!t]
\centering
\resizebox{0.95\textwidth}{!}{% =====================================================================
%  Truncation strategies (Fig. 1). Preamble-free TikZ body, \input by
%  sections/02-mathematical_bg-2.tex so that its symbols and fonts are
%  typeset exactly as in the text. Standalone preview:
%  TikzDiagrams/truncation.tex, compiled from TikzDiagrams/.
%  Uses \T, \V, \Rn, \Tra from macros/macros.sty.
% =====================================================================
% ============================================================
% Projection
% ============================================================
\providecommand{\tpoint}[3]{({#1 + 0.5*#3},{#2 + 0.4*#3})}

% ============================================================
% Slice drawing macro
% ============================================================

\providecommand{\drawslice}[6]{%
    \pgfmathsetmacro{\xxA}{#2}
    \pgfmathsetmacro{\xxB}{#2 + #4}
    \pgfmathsetmacro{\yyA}{#3}
    \pgfmathsetmacro{\yyB}{#3 + #5}

    \filldraw[
        fill=#6!65,
        fill opacity=0.35,
        draw=black!75,
        draw opacity=0.90,
        line width=0.30pt
    ]
        \tpoint{\xxA}{\yyA}{#1}
        --
        \tpoint{\xxB}{\yyA}{#1}
        --
        \tpoint{\xxB}{\yyB}{#1}
        --
        \tpoint{\xxA}{\yyB}{#1}
        -- cycle;
}

% ============================================================
% Bounding box macro
% ============================================================

\providecommand{\drawtensorbox}[3]{%

    % back face
    \draw[black!70]
        \tpoint{0}{0}{#3}
        --
        \tpoint{#1}{0}{#3}
        --
        \tpoint{#1}{#2}{#3}
        --
        \tpoint{0}{#2}{#3}
        -- cycle;

    % front face
    \draw[black!70]
        \tpoint{0}{0}{0}
        --
        \tpoint{#1}{0}{0}
        --
        \tpoint{#1}{#2}{0}
        --
        \tpoint{0}{#2}{0}
        -- cycle;

    % connectors
    \draw[black!70]
        \tpoint{0}{0}{0}
        --
        \tpoint{0}{0}{#3};

    \draw[black!70]
        \tpoint{#1}{0}{0}
        --
        \tpoint{#1}{0}{#3};

    \draw[black!70]
        \tpoint{#1}{#2}{0}
        --
        \tpoint{#1}{#2}{#3};

    \draw[black!70]
        \tpoint{0}{#2}{0}
        --
        \tpoint{0}{#2}{#3};
}

\begin{tikzpicture}[
    scale=0.92,
    line join=round,
    line cap=round,
    >=Stealth
]

% ============================================================
% Parameters
% ============================================================

\def\sliceSep{0.32}
\def\depth{5*\sliceSep}

% Full tensor dimensions
\def\mDim{2.4}
\def\pDim{1.8}

% Truncated rank
\def\rDim{1.25}

% Variable rank max
\def\maxrho{1.35}

% Vertical shift for SVT
\pgfmathsetmacro{\svShift}{\mDim-\rDim}

% Small correction for t-SVDM-II SVT
\def\svAdjust{-0.10}

% Randomized variable ranks
\def\rA{1.18}
\def\rB{0.42}
\def\rC{1.34}
\def\rD{0.73}
\def\rE{0.91}
\def\rF{0.56}

% Common n-arrow x position
\def\nArrowX{2.12}

% ============================================================
% CENTER : FULL TENSOR
% ============================================================

\begin{scope}[xshift=10cm]

\foreach \z in {5,4,3,2,1,0} {
    \drawslice{\z*\sliceSep}{0}{0}{\pDim}{\mDim}{green}
}

\drawtensorbox{\pDim}{\mDim}{\depth}

% Titles
\node[font=\normalsize\bfseries]
    at (1.5,4.15)
    {Full Tensor};

\node[font=\small]
    at (1.55,-1.2)
    {$\T[\widehat]{A}\in\Rn{d_1\times d_2\times d_3}$};

% m
\draw[<->]
    (-0.35,0)
    --
    (-0.35,\mDim)
    node[midway,left,font=\footnotesize] {$d_1$};

% p
\draw[<->]
    (0,-0.30)
    --
    (\pDim,-0.30)
    node[midway,below,font=\footnotesize] {$d_2$};

% n
\draw[<->,gray]
    \tpoint{\nArrowX}{\mDim}{\depth}
    --
    \tpoint{\nArrowX}{\mDim}{0};

\node[gray,font=\footnotesize]
    at \tpoint{\nArrowX+0.28}{\mDim}{0.5*\depth}
    {$d_3$};

\end{scope}

% ============================================================
% LEFT ARROW TO t-SVDM
% ============================================================

\draw[
    line width=1.4pt,
    -{Stealth[length=5mm,width=3mm]},
    shorten >=2pt
]
    (9.1,1.4)
    --
    (7.2,1.4);

% ============================================================
% RIGHT ARROW TO t-SVDM-II
% ============================================================

\draw[
    line width=1.4pt,
    -{Stealth[length=5mm,width=3mm]},
    shorten >=2pt
]
    (12.9,1.4)
    --
    (14.8,1.4);

% ============================================================
% LEFT : t-SVDM
% ============================================================

\begin{scope}[xshift=0cm]

% ------------------------------------------------------------
% U tensor
% ------------------------------------------------------------

\begin{scope}

\foreach \z in {5,4,3,2,1,0} {
    \drawslice{\z*\sliceSep}{0}{0}{\rDim}{\mDim}{cyan}
}

\drawtensorbox{\pDim}{\mDim}{\depth}

\node[font=\small]
    at (1.05,-1.20)
    {$\T[\widehat]{U}_k\in\Rn{d_1\times k\times d_3}$};

% m
\draw[<->]
    (-0.35,0)
    --
    (-0.35,\mDim)
    node[midway,left,font=\footnotesize] {$d_1$};

% r
\draw[<->]
    (0,-0.30)
    --
    (\rDim,-0.30)
    node[midway,below,font=\footnotesize] {$k$};

% n
\draw[<->,gray]
    \tpoint{\nArrowX}{\mDim}{\depth}
    --
    \tpoint{\nArrowX}{\mDim}{0};

\node[gray,font=\footnotesize]
    at \tpoint{\nArrowX+0.28}{\mDim}{0.5*\depth}
    {$d_3$};

\end{scope}

% ------------------------------------------------------------
% SVT tensor
% ------------------------------------------------------------

\begin{scope}[xshift=4.0cm,yshift=\svShift cm]

\foreach \z in {5,4,3,2,1,0} {
    \drawslice{\z*\sliceSep}{0}{0}{\pDim}{\rDim}{magenta}
}

\begin{scope}[yshift=-\svShift cm]
    \drawtensorbox{\pDim}{\mDim}{\depth}
\end{scope}

\node[font=\small]
    at (1.05,-2.32)
    {$\T[\widehat]{S}_k\triangle\T[\widehat]{V}_k^{\Tra}\in\Rn{k\times d_2\times d_3}$};

% k
\draw[<->]
    (-0.35,0)
    --
    (-0.35,\rDim)
    node[midway,left,font=\footnotesize] {$k$};

% p
\draw[<->]
    (0,-1.45)
    --
    (\pDim,-1.45)
    node[midway,below,font=\footnotesize] {$d_2$};

\end{scope}

% titles
\node[font=\normalsize\bfseries]
    at (3.45,4.15)
    {t-SVDM-I};

\node[font=\small]
    at (3.45,3.55)
    {};

\end{scope}

% ============================================================
% RIGHT : t-SVDM-II
% ============================================================

\begin{scope}[xshift=15.8cm]

% ------------------------------------------------------------
% U tensor
% ------------------------------------------------------------

\begin{scope}

\foreach \rho/\z in {
    \rE/5,
    \rD/4,
    \rC/3,
    \rB/2,
    \rA/1,
    \rF/0
} {
    \drawslice{\z*\sliceSep}{0}{0}{\rho}{\mDim}{cyan}
}

\drawtensorbox{\pDim}{\mDim}{\depth}

\node[font=\small]
    at (1.05,-1.20)
    {$\T[\widehat]{U}_{\V{\rho}}$ (multi-rank $\V{\rho}$)};

% m
\draw[<->]
    (-0.35,0)
    --
    (-0.35,\mDim)
    node[midway,left,font=\footnotesize] {$d_1$};

% rho1
\draw[<->]
    (0,-0.30)
    --
    (\rF,-0.30)
    node[midway,below,font=\footnotesize] {$\rho_1$};

% rho_n
\draw[<->]
    (0.7,3.18)
    --
    (0.85 + \rE,3.18);

\node[font=\small]
    at (1.2,3.40)
    {$\rho_{d_3}$};

% n
\draw[<->,gray]
    \tpoint{\nArrowX}{\mDim}{\depth}
    --
    \tpoint{\nArrowX}{\mDim}{0};

\node[gray,font=\footnotesize]
    at \tpoint{\nArrowX+0.28}{\mDim}{0.5*\depth}
    {$d_3$};

\end{scope}

% ------------------------------------------------------------
% SVT tensor
% ------------------------------------------------------------

\begin{scope}[xshift=4.0cm,yshift=\svShift cm]

\begin{scope}[yshift=\svAdjust cm]

\foreach \rho/\z in {
    \rE/5,
    \rD/4,
    \rC/3,
    \rB/2,
    \rA/1,
    \rF/0
} {
    \drawslice{\z*\sliceSep}{0}{\maxrho-\rho}{\pDim}{\rho}{magenta}
}

% rho1
\draw[<->]
    (-0.35,\maxrho-\rF)
    --
    (-0.35,\maxrho)
    node[midway,left,font=\footnotesize] {$\rho_1$};

% rho_n
\draw[<->]
    (\pDim+1.02,\maxrho-\rE + 0.7)
    --
    (\pDim+1.02,\maxrho + 0.7);

\node[font=\small]
    at (\pDim+1.42,\maxrho+0.3)
    {$\rho_{d_3}$};

% p
\draw[<->]
    (0,-1.45)
    --
    (\pDim,-1.45)
    node[midway,below,font=\footnotesize] {$d_2$};

\end{scope}

% bounding box
\begin{scope}[yshift=-\svShift cm]
    \drawtensorbox{\pDim}{\mDim}{\depth}
\end{scope}

\node[font=\small]
    at (1.05,-2.30)
    {$\T[\widehat]{S}_{\V{\rho}}\triangle\T[\widehat]{V}_{\V{\rho}}^{\Tra}$ (multi-rank $\V{\rho}$)};

\end{scope}

% titles
\node[font=\normalsize\bfseries]
    at (3.45,4.15)
    {t-SVDM-II};

\node[font=\small]
    at (3.45,3.55)
    {};

\end{scope}

\end{tikzpicture}}
\caption{
Illustration of tensor truncation strategies.
The center tensor represents the full transformed tensor.
\tsvdmi applies a uniform truncation rank across all transformed frontal slices,
whereas \tsvdmii determines adaptive per-slice truncation ranks based on the spectral energy distribution.
}
\label{fig:tsvdm_truncation}
\end{figure*}
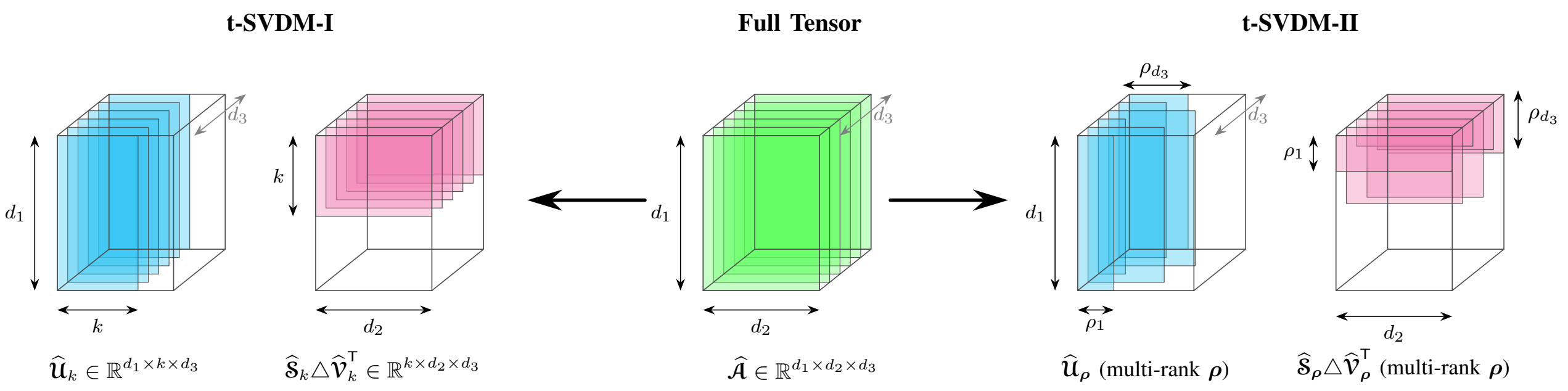

If we instead allow different truncation levels to vary across the frontal slices, say $\rho_{f_i}$ for face $f_i$, we can obtain an adaptive variant of the t-SVDM. $\V{\rho}$ is called the \emph{multi-rank} of $\T{A}$.

\paragraph{t-SVDM-II} The t-SVDM-II decomposition is the adaptive, multi-rank variant of t-SVDM for $N$-way tensors. As illustrated in \Cref{fig:tsvdm_truncation}, instead of enforcing a single global rank $k$ across all transform-domain frontal slices, t-SVDM-II truncates each slice according to its own target rank $\rho_{f_i}$. Equivalently, after computing slice-wise SVDs of $\T[\widehat]{A}$, one retains the first $\rho_{f_i}$ singular triplets in each slice $f_i$, producing factors that reconstruct the multi-rank approximation
\begin{equation}
\T{A}_{\V{\rho}} = \T{U}_{\V{\rho}} \starM \T{S}_{\V{\rho}} \starM \T{V}_{\V{\rho}}^{\Tra}.
\end{equation}
This adaptive truncation provides finer control of the compression-accuracy trade-off when spectral decay varies across slices.

\begin{algorithm}[hbtp]
\caption{$N$-Way t-SVDM-II with tolerance $\epsilon$}
\label{alg:tsvdmii}
\begin{algorithmic}
\REQUIRE Input tensor $\T{A} \in \Rn{d_1 \times \cdots \times d_N}$, transformation matrices $\Curly{\M{M}_i \in \Rn{d_i \times d_i}}_{i=3}^N$, and tolerance $\epsilon$.
\ENSURE $\pr{\T{U}_{\V{\rho}}, \T{S}_{\V{\rho}}, \T{V}_{\V{\rho}}}$ with multi-rank ${\V{\rho}}$ such that $\norm{\T{A}-\T{A}_{\V{\rho}}}_F \le \epsilon\norm{\T{A}}_F$.
\STATE {\textsc{t-SVDM-II}}$(\T{A}, \Curly{\M{M}_i}, \epsilon)$
\STATE $\T[\widehat]{A} \gets \T{A} \times_3 \mathbf{M}_3 \times_4 \cdots \times_N \mathbf{M}_N $
\FOR{$f_i \in F$}
%\STATE $\T[\widehat]{U}_{f_i}, \T[\widehat]{S}_{f_i}, \T[\widehat]{V}_{f_i} \gets \textsc{SVD}\pr{\M[\widehat]{A}_{f_i}}$
\STATE $\M[\widehat]{U}_{f_i}, \M[\widehat]{S}_{f_i}, \M[\widehat]{V}_{f_i} \gets \textsc{SVD}\pr{\M[\widehat]{A}_{f_i}}$
\ENDFOR
\STATE $\V{v} \gets \textsc{concatenate}\pr{\mathrm{diag}\pr{\M[\widehat]{S}_{f_i}}^2} \forall f_i \in F$.
\STATE $\V{v} \gets \textsc{sort}(\V{v},\text{ascending})$
\STATE $w_k = \sum_{\ell=1}^{k} v_\ell$
\STATE Find the largest index $J$ such that
$\frac{w_J}{\norm{\T[\widehat]{S}}_F^2}
<
\epsilon^2
$
\STATE  $\tau \gets v_J$
\FOR{$f_i \in F$}
\STATE $\rho_{f_i} \gets 
\#\left\{
\M[\widehat]{S}_{f_i}
>
\tau
\right\}$
%\STATE $\rho_{f_i} = k$
%\STATE $\T[\widehat]{U}_{f_i}, \T[\widehat]{S}_{f_i}, \T[\widehat]{V}_{f_i} \gets \M[\widehat]{U}_{f_i}(:,1:k), \M[\widehat]{S}_{f_i}(1:k,1:k), \M[\widehat]{V}_{f_i}(:,1:k)$
\STATE $\T[\widehat]{U}_{\V{\rho}}^{f_i}, \T[\widehat]{S}_{\V{\rho}}^{f_i}, \T[\widehat]{V}_{\V{\rho}}^{f_i} \gets \textsc{truncate}\pr{\M[\widehat]{U}_{f_i}, \M[\widehat]{S}_{f_i}, \M[\widehat]{V}_{f_i}, \rho_{f_i}}$
\ENDFOR
\STATE $\T{U}_\rho \gets \T[\widehat]{U}_\rho \times_3 \mathbf{M}_3^{-1} \times_4 \cdots \times_N \mathbf{M}_N^{-1}$
\STATE $\T{S}_\rho \gets \T[\widehat]{S}_\rho \times_3 \mathbf{M}_3^{-1} \times_4 \cdots \times_N \mathbf{M}_N^{-1}$
\STATE $\T{V}_\rho \gets \T[\widehat]{V}_\rho \times_3 \mathbf{M}_3^{-1} \times_4 \cdots \times_N \mathbf{M}_N^{-1}$
\RETURN $\pr{\T{U}_\rho, \T{S}_\rho, \T{V}_\rho}$
\end{algorithmic}
\end{algorithm}

\paragraph{Optimality}
A key advantage of the t-SVDM framework is that it admits Eckart--Young--Mirsky-type optimality guarantees. Kilmer et al. \cite{kilmer2021tensor} prove that when the transformation matrices are nonzero multiples of unitary matrices, both the t-SVDM-I and t-SVDM-II truncations are \emph{exactly} optimal in terms of Frobenius norm minimization over the set of all t-rank or multi-rank constrained tensors. %\TODO{Srini, for the face use superscript instead of subscript and write Higher Order tensor SVD in section instead of Higher order SVD.} Contrast this with other popular tensor decompositions like CP, Tucker, HOSVD or Tensor-Train where no such global guarantees exist \cite{kolda2009tensor}. Only \emph{quasi-optimal} solutions can be found for HOSVD, Tucker, and Tensor-Train decompositions.

\subsection{Relevant Tensor Libraries}
%\begin{enumerate}
%	\item \texttt{pystarm} and lack of HPC $\starM$ libraries.
%	\item Distributed-memory libraries for CP/Tucker and maybe other decompositions.
%	\item Differentiation of our work.
%\end{enumerate}
Despite the mathematically pleasing features of the t-SVDM algorithm, very few high-performance implementations of these methods exist. Most of the original research and reference implementations were in productivity-oriented languages like Matlab and Python \cite{kilmer2011factorization,kilmer2021tensor,newman2019step}. PyStarM provided the first shared-memory parallel implementation, using threaded batched-BLAS and LAPACK kernels, of both t-SVDM variants \cite{hussain2026highperformancestarmsvdbig}. The authors analyzed the two main phases on the algorithms, TTM and SVD operations, and discussed various design choices. 

Distributed-memory parallel versions of other tensor decompositions have a more mature history. CP libraries are the most popular with many specialized to incorporate sparse tensors, constraints on factors, and hardware accelerators \cite{phipps2017genten,phipps2019software,smith2016medium,smith2015splatt,eswar2021planc}. Tucker/HOSVD \cite{ballard2020tuckermpi,chakaravarthy2017optimizing,chakaravarthy2018optimizing}, Tensor-Train \cite{daas2022parallel,shi2023parallel}, and tensor networks \cite{solomonik2013cyclops} are other decompositions with distributed-memory libraries.

In this work, we consider the first distributed-memory implementation of the t-SVDM algorithms. Besides the algorithmic contributions, the ability to scale to multiple nodes will provide another avenue for practitioners to analyze data within the $\starM$ framework and enjoy the optimality guarantees that come with it.

% \input{sections/03-methodology.tex}
%!TEX root = ../main.tex

% New methodology section, drafted alongside sections/03-methodology.tex.
% Both are included from main.tex; this one uses \label{sec:methodology-new}
% so it does not collide with \label{sec:methodology} in the old section.

\section{Methodology}\label{sec:methodology-new}

\begin{figure*}[t]
  \centering
  \input{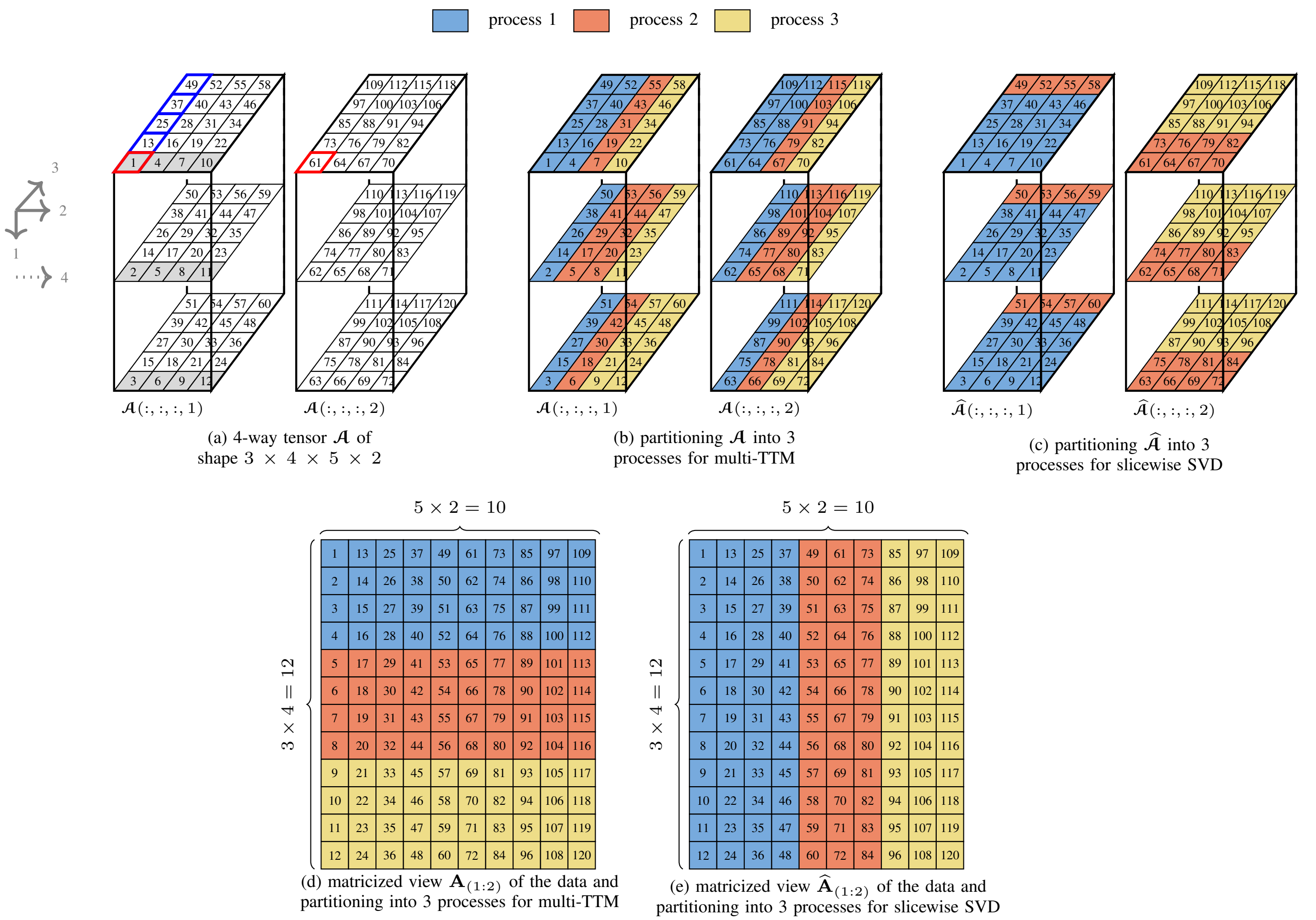}
  \caption{A 4-way tensor $\T{A}$ and its distribution among 3 processors. Blue
  highlighted cells in (a) represent one mode-3 fiber and red highlighted cells
  in (a) represent one mode-4 fiber. Gray highlighted cells in (a) represent one
  frontal slice of the tensor.}
  \label{fig:distribution}
\end{figure*}

\begin{figure}[t]
	\centering
	\input{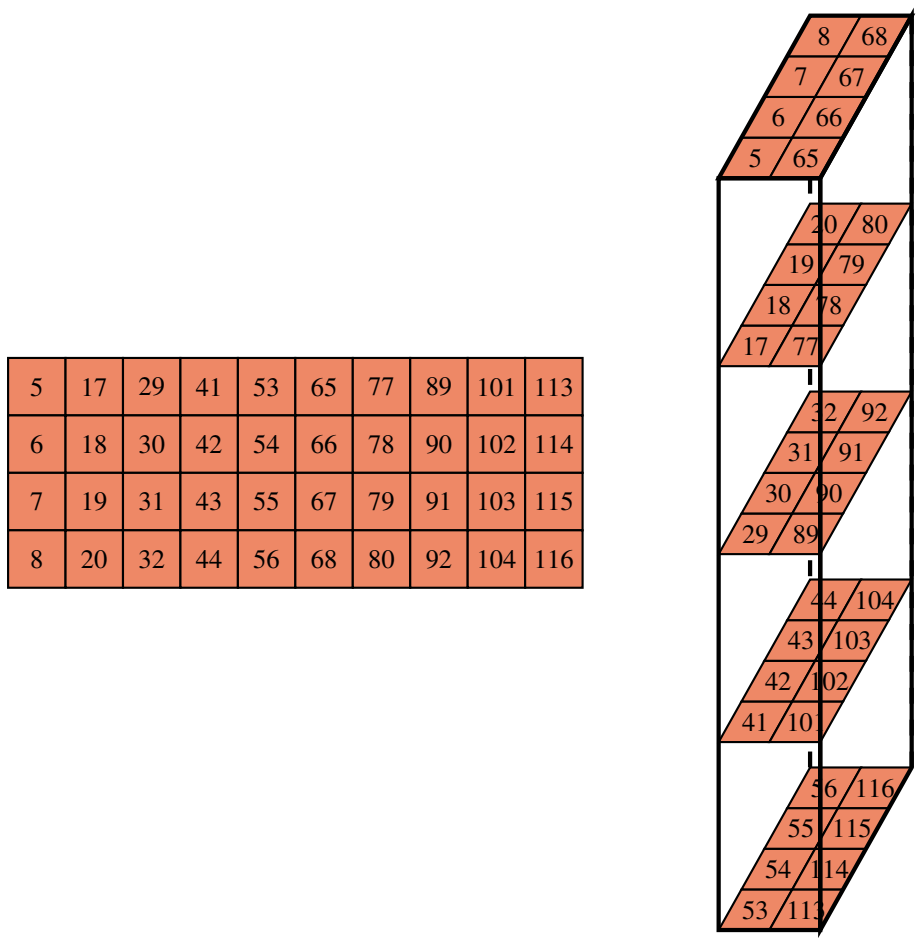}
	\caption{Local data $\underline{\Mz{A}{1:2}}^{[2]}$ of process 2 in the TTM phase, for the same
	$3\times4\times5\times2$ tensor and 3-processor distribution shown in
	\cref{fig:distribution}. The partitioning turns the local part of the
	4-way tensor into a 3-way local tensor, and the multi-TTM is performed
	locally on all of its modes but one.}
	\label{fig:ttm_local_p2}
\end{figure}

In this section we describe our MPI parallelization of \tsvdmi and
\tsvdmii for $N$-way tensors. Both algorithms have two phases. The
first is a multi-TTM, and the second is a slicewise SVD.  
Note that our algorithms return the t-SVD factors in the transformed domain; in order to produce the factors in the original domain as presented in \cref{alg:tsvdmi,alg:tsvdmii}, another set of multi-TTMs would be required.

\Cref{fig:distribution} motivates and illustrates our overall approach for a small 4-way example tensor.
As shown in \cref{fig:distribution}(a) the multi-TTM updates fibers in modes 3 and 4, and an example of each are highlighted in blue and red, respectively.
Given the transformation matrices $\mathbf{M}_3$ and $\mathbf{M}_4$, each fiber can be updated independently and in parallel.
The slicewise SVDs are performed on frontal slices, which corresponds to modes 1 and 2, and an example frontal slice is highlighted in gray.
Each slice can be decomposed independently and in parallel.

Our approach seeks to capitalize on the independence within each of the phases, but we observe that each phase calls for a different data distribution.
We therefore use one distribution per phase and redistribute the tensor in between.
As we show in \cref{fig:distribution} and describe in more detail in \cref{sec:partitioning}, we choose a distribution in the multi-TTM phase that ensures that each fiber in modes 3 through $N$ is local to a single processor, and we choose a distribution in the slicewise SVD phase that ensures that each frontal slice is local to a single processor.

%Running the two phases with minimum communication requires two
%different data layouts. A TTM along mode $k$ transforms every mode-$k$
%fiber of the tensor. t-SVDM applies a TTM to modes $3$ through $N$. So
%every fiber is transformed except the mode-1 and mode-2 fibers. For
%the multi-TTM phase, each process should therefore own whole fibers
%along modes $3$ to $N$. How we split modes 1 and 2 does not matter
%here. The SVD phase treats each frontal slice as a matrix. For this
%phase each process should own whole frontal slices, that is, whole
%mode-1 and mode-2 fibers. How we split the other modes does not matter
%here. The two requirements conflict. We therefore use one layout per
%phase and redistribute the tensor in between.

The distribution chosen for multi-TTM leads to a communication-optimal 
1D matrix multiplication algorithm for each TTM assuming $p \le \mathcal{N}/ d_k^2$  for each $3\le k\le N$, where $\mathcal{N} = \prod_{\ell=1}^N d_\ell$ and $p$ is the number of processors \cite{ABGKR22}.
%$d_k^2 \leq \mathcal{N}/p$
In our approach, the only communication required of the multi-TTM is the sharing of the transformation matrices across all processors, which are typically small relative to the input tensor.
If the transformation can be generated locally, as in the case of a discrete cosine transform, then no communication is required.
The distribution chosen for the slicewise SVDs avoids all communication as well in the case of \tsvdmi.
In the case of \tsvdmii, truncation ranks are determined from global singular value information, and the communication performed in this step is described in \cref{sec:truncation}.
The main communication cost for both algorithms is that of the tensor redistribution. 
We leave the question of communication optimality of the entire \tsvdmi and \tsvdmii algorithms to future work.

%Two properties of the problem shape this design. The first concerns
%the transformation matrices. Each one is square with the size of a
%single mode, so it is normally small next to the tensor, and every
%process can generate and keep its own copy. This is not true in every
%case. A three-way tensor whose first two modes are tiny and whose
%third mode is large has a transformation matrix bigger than the tensor
%itself, and replicating it would then be the wrong choice. Such shapes
%are rare in practice, so we assume throughout that the transformation
%matrices are small compared to the tensor.
%
%The second concerns where the time goes. The slicewise SVDs dominate
%the runtime for practical datasets, while the multi-TTM is
%comparatively cheap. One could therefore keep a single layout chosen
%to suit the SVD phase, and accept communication inside the multi-TTM
%instead. But a distributed multi-TTM has to move at least one copy of
%the data in any case. Paying a single redistribution and leaving both
%compute phases free of communication is therefore no worse in
%practice. We take that route: both compute phases stay local to a
%process, and all communication is confined to the redistribution
%between them.

%\subsection{Data Partitioning}
\subsection{Parallelization and Data Partitioning}
\label{sec:partitioning}

\subsubsection{Data Partitioning for TTM phase}
To distribute the tensor for the multi-TTM phase, we first conceptually linearize
the first two modes. This transforms the $N$-way tensor into an $(N-1)$-way
tensor whose first mode runs over all pairs of the first two indices.
We then divide that first mode evenly among the processors. 
This is not a usual block distribution of the original tensor, but it is a block
distribution of the linearized tensor onto a processor grid of shape
$p \times 1 \times \cdots \times 1$. 
\Cref{fig:distribution}(a) shows a 4-way example tensor of shape $3 \times 4 \times 5 \times 2$, and \cref{fig:distribution}(b) shows how we partition it across three processors.

This layout keeps every transformed fiber local to a single processor. 
A TTM on modes $3$ through $N$ transforms fibers that lie entirely inside one subtensor, and each subtensor is local to a single process. 
In \cref{fig:distribution}(a), the blue cells are a mode-3 fiber and the red cells are a mode-4 fiber. 
Both lie inside one local subtensor, so both can be transformed with no communication.

The same layout can be illustrated in a matricized view. 
We reshape the tensor in \cref{fig:distribution}(b)  into a matrix in \cref{fig:distribution}(d) whose rows are indexed by the first two modes and
whose columns are indexed by the remaining modes. 
Both linearizations
follow natural order (\Cref{sec:notation}). 
In this view our partitioning is a block row partitioning, where each processor owns a contiguous block of rows. 
	
\subsubsection{Local TTM}
After the partitioning above, the multi-TTM needs no communication at
all. Every processor transforms only the fibers it owns. The
transformation matrices are small compared to the tensor, so every
processor keeps its own copy of each one.

Each processor stores its local data as an $(N-1)$-way tensor.
Its first $N-2$ modes are modes $3$ through $N$ of the original
tensor, and its last mode corresponds to the linearization of the original first two modes. 
Permuting the linearized first two modes to the last helps local performance of the TTM in the original mode 3.
In the example of \cref{fig:distribution}, this local tensor has shape $5 \times 2 \times 4$.

\Cref{fig:ttm_local_p2}(a) shows the local block of the second process
for our example tensor, taken from the matricized view of
\cref{fig:distribution}(d). Reading those entries in row-major order
gives the 3-way tensor of \cref{fig:ttm_local_p2}(b), whose modes are
the original modes 3 and 4 followed by the trailing linearized mode.

We then apply one TTM for each of modes $3$ through $N$ of the
original tensor and skip the trailing mode. 
In \cref{fig:ttm_local_p2}(b) this means
transforming the first two modes and leaving the third alone. Each TTM
is a single (batched) BLAS call involving the whole local tensor, using the optimized local TTM algorithms described in
\cite{Ballard_Kolda_2025,hussain2026highperformancestarmsvdbig}.

\subsubsection{Data Partitioning for SVD phase}
The SVD phase treats each frontal slice as a matrix and factors it.
Each factorization needs the whole slice, which is local to a single process. 
\Cref{fig:distribution}(a) shows one frontal slice
of the example tensor in gray.

To partition for this phase we conceptually linearize modes $3$ through $N$ and
divide this 3rd mode evenly among the processors. When the count
does not divide evenly, some processors take one slice more than the
others. Our example has ten frontal slices and three processors, so
they split as four, three and three. \Cref{fig:distribution}(c) shows
this partitioning.

The matricized view is the same one we used for the TTM phase. There
we cut the matrix into blocks of rows. Here we cut it into blocks of
columns, so each processor owns a contiguous block of columns.
\Cref{fig:distribution}(e) shows the matricized view of the same
partitioning as \cref{fig:distribution}(c).

\begin{figure}[t]
	\centering
	\input{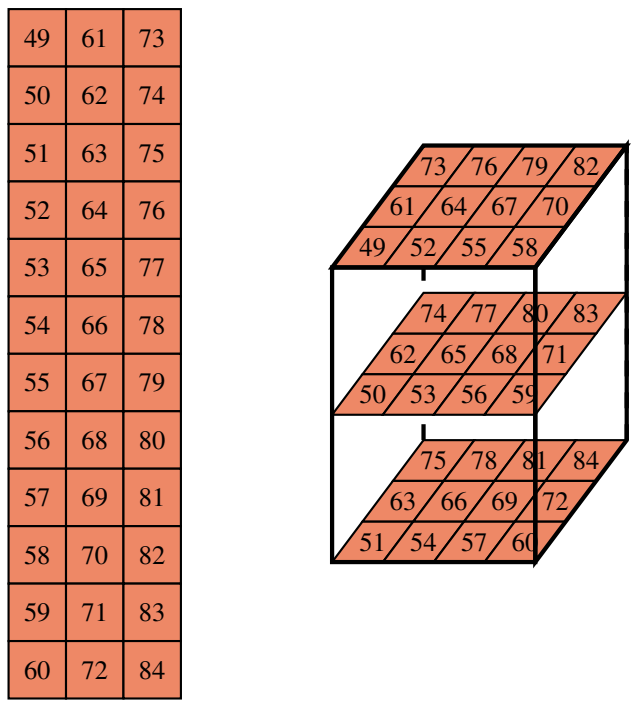}
	\caption{Local data $\underline{\Mz[\widehat]{A}{1:2}}^{[2]}$ of process 2 in the SVD phase, for the same
	$3\times4\times5\times2$ tensor and 3-processor distribution shown in
	\cref{fig:distribution}. The partitioning turns the local part of the
	4-way tensor into a 3-way local tensor, and each of its frontal slices is
	factored locally by an SVD.}
	\label{fig:svd_local_p2}
\end{figure}

\subsubsection{Local SVD}
Partitioning for the SVD phase gives each processor a set of 
frontal slices. A processor holds every entry of the slices it owns, so
it can factor them without any further communication.

\Cref{fig:svd_local_p2}(a) shows the local block of the second processor,
the columns of matrix view shown in \cref{fig:distribution}(e) that it owns. These entries
are contiguous in memory, and reading them in column-major order gives
the 3-way tensor of \cref{fig:svd_local_p2}(b). Its first two modes are
modes 1 and 2 of the original tensor, and its last mode is a local part of the linearization of all other original modes. 
Each frontal slice of this tensor is a complete frontal slice of the original tensor, that is, one of the matrices the SVD factors.

We factor each local frontal slice with an optimized sequential SVD and
apply the truncation. The slices carry no data dependency between them,
so every processor works through its own slices independently. We overlap
these factorizations with the redistribution that precedes them, which we
describe next.

\subsection{Overlapping the SVD Phase with Redistribution}
\label{sec:alg:overlap}

The computational cost is typically dominated by the slicewise SVD phase, and the communication cost is the redistribution of tensor data between phases.
We now describe how we hide the communication cost by overlapping these two operations.

As illustrated in \cref{fig:distribution}, the redistribution involves each processor exchanging approximately $\mathcal{N}/p^2$ data with every other processor.
We can perform the redistribution with a single all-to-all exchange between the two phases. 
Almost every entry of the tensor changes processors, so the bandwidth cost (amount of data communicated by each processor) of the exchange is $\Omega(\mathcal{N}/p)$.

%We accept this cost because of what it trades off. The alternative is to
%leave the tensor in the TTM layout and factor each frontal slice with
%a distributed SVD. That spreads every slice across many processors and
%puts communication inside every factorization. A single exchange
%instead pays for communication once and leaves every SVD local.

Much of that one-time cost can be hidden. The slicewise SVDs are
independent of one another, so a processor does not have to wait for the
whole exchange to finish before it starts its computation. As soon as one complete frontal slice has
arrived, the processor can start factoring it.

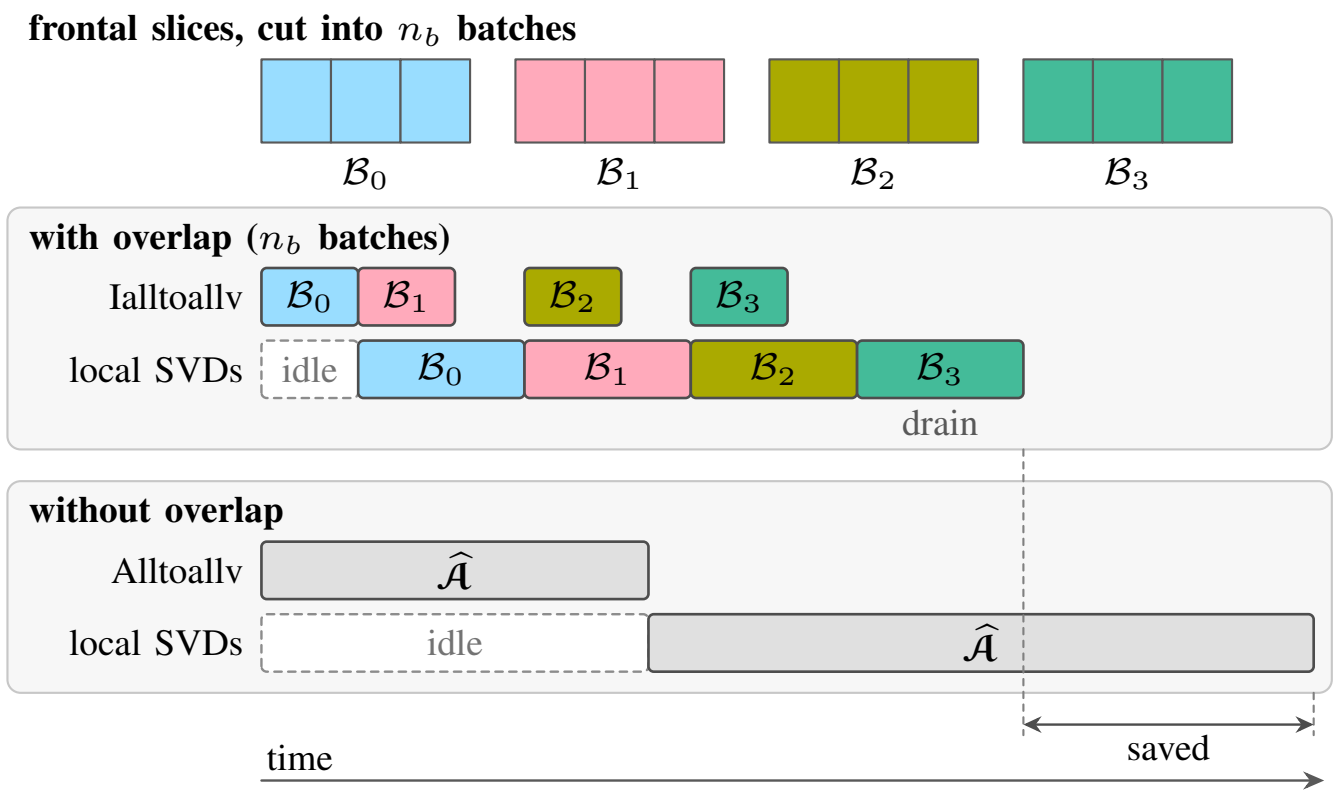
\begin{figure}[t]
  \centering
  \providecommand{\T}[2][]{\boldsymbol{#1{\mathscr{\MakeUppercase{#2}}}}}%
\begin{tikzpicture}[
  line join=round, line cap=round, >=Stealth, font=\footnotesize,
  slice/.style={draw=black!65, line width=0.4pt},
  block/.style={rounded corners=1pt, draw=black!70, line width=0.5pt},
  card/.style={rounded corners=2.5pt, draw=black!22, fill=black!3,
               line width=0.4pt},
  idlebox/.style={rounded corners=1pt, draw=black!45, fill=white,
                  dash pattern=on 1.4pt off 1.2pt, line width=0.45pt},
  % all four lane labels use the body font, not \texttt, so the gutter
  % reads uniformly; the MPI calls are marked up properly in the caption
  lanelab/.style={font=\scriptsize, anchor=east, inner sep=1.5pt},
  grouplab/.style={font=\scriptsize\bfseries, anchor=north west, inner sep=1pt},
  blab/.style={font=\scriptsize, inner sep=0.8pt},
  ann/.style={font=\scriptsize, inner sep=0.8pt},
  guide/.style={dash pattern=on 1.2pt off 1.2pt, draw=black!50,
                line width=0.4pt},
  meas/.style={{Stealth[length=1.2mm,width=1.1mm]}-{Stealth[length=1.2mm,width=1.1mm]},
               draw=black!70, line width=0.45pt},
]

% ------------------------------------------------------------------
% Knobs
% ------------------------------------------------------------------
% Four batches at \scriptsize is a tight fit: the sequential card needs
% n_b*(cc+ss) to sit inside one column, and the gutter holding the lane
% labels costs a fixed ~1.6cm on top of that.  What buys the room is
% keeping "drain" OUT of the last SVD block -- as a block label it forces
% ss wide enough to hold "B_3 (drain)", which four batches cannot afford.
% It is written under that block instead, where it reads the same.
\def\NB{4}             % number of batches n_b
\def\SPB{3}            % frontal slices drawn per batch
\def\cw{0.46}          % slice cell width
\def\bgap{0.30}        % gap between batches
\def\t0{1.20}          % time origin (left edge of every lane)
\def\cc{0.64}          % duration of one batch exchange (must hold "idle")
\def\ss{1.10}          % duration of one batch of local SVDs
\def\lh{0.19}          % half lane height
% Card left edge.  The lane labels are the leftmost objects in the
% picture: at \scriptsize they reach x = -0.33, so this has to sit left of
% that or the "l" of "local SVDs" hangs outside the card.  Raising the
% lane-label font means moving this too.
\def\cardL{-0.48}      % card left edge  (padding for the lane labels)
\def\cardR{8.28}       % card right edge
\def\titleX{0.10}      % every group title sits at \cardL+\titleX
% Each group title clears the content below it by ~0.15cm; the y values
% below already include that gap, so shifting a lane means shifting its
% title too.
\def\yA{-0.635}        % slice-strip centre
\def\yCa{-1.93}        % overlapped card, communication lane
\def\ySa{-2.41}        % overlapped card, computation lane
\def\yCb{-3.74}        % sequential card, communication lane
\def\ySb{-4.22}        % sequential card, computation lane
\pgfmathsetmacro{\NBmax}{\NB-1}   % last batch index, for the \foreach loops
\pgfmathsetmacro{\bw}{\SPB*\cw}
\pgfmathsetmacro{\pitch}{\bw+\bgap}
\pgfmathsetmacro{\nbA}{\t0+\NB*\cc}          % end of the single exchange
\pgfmathsetmacro{\tEnd}{\t0+\cc+\NB*\ss}     % overlapped finish
\pgfmathsetmacro{\bEnd}{\t0+\NB*\cc+\NB*\ss} % non-overlapped finish

% ---- Paul Tol "light" qualitative palette (SRON), the same one the tensor
% and distribution figures use: safe for deuteranopia, protanopia and
% tritanopia, and light enough to keep black labels legible on top.
% https://sronpersonalpages.nl/~pault/
\definecolor{tolBlue}{HTML}{77AADD}%
\definecolor{tolRed}{HTML}{EE8866}%
\definecolor{tolYellow}{HTML}{EEDD88}%
\definecolor{tolMint}{HTML}{44BB99}%
\definecolor{tolLightCyan}{HTML}{99DDFF}%
\definecolor{tolPink}{HTML}{FFAABB}%
\definecolor{tolOlive}{HTML}{AAAA00}%
\def\bfill#1{\ifcase#1 tolLightCyan\or tolPink\or tolOlive\or tolMint\fi}

% a timeline block: #1 x1, #2 x2, #3 y centre, #4 fill, #5 label
\def\tblock#1#2#3#4#5{%
  \draw[block, fill=#4] (#1,{#3-\lh}) rectangle (#2,{#3+\lh});
  \node[blab] at ({0.5*(#1+#2)},#3) {#5};%
}
% a stall in a lane: same signature, no fill colour
\def\idle#1#2#3#4{%
  \draw[idlebox] (#1,{#3-\lh}) rectangle (#2,{#3+\lh});
  \node[blab, text=black!55] at ({0.5*(#1+#2)},#3) {#4};%
}

% ==================================================================
% how the slices are batched
% ==================================================================
\node[grouplab] at ({\cardL+\titleX},-0.02)
  {frontal slices, cut into $n_b$ batches};
\foreach \b in {0,...,\NBmax}{%
  \pgfmathsetmacro{\bx}{\t0+\b*\pitch}
  \foreach \i in {0,...,2}{%
    \pgfmathsetmacro{\cx}{\bx+\i*\cw}
    \draw[slice, fill=\bfill{\b}] (\cx,-0.36) rectangle ({\cx+\cw},-0.91);
  }
  \node[ann, anchor=north] at ({\bx+0.5*\bw},-0.98) {$\mathcal{B}_\b$};
}

% ==================================================================
% card 1: with overlap
% ==================================================================
\draw[card] (\cardL,-1.34) rectangle (\cardR,-2.94);
\node[grouplab] at ({\cardL+\titleX},-1.40) {with overlap ($n_b$ batches)};
\node[lanelab] at ({\t0-0.06},\yCa) {Ialltoallv};
\node[lanelab] at ({\t0-0.06},\ySa) {local SVDs};

\foreach \b in {0,...,\NBmax}{%
  \ifnum\b=0\relax
    \pgfmathsetmacro{\xa}{\t0}
  \else
    \pgfmathsetmacro{\xa}{\t0+\cc+(\b-1)*\ss}
  \fi
  \tblock{\xa}{\xa+\cc}{\yCa}{\bfill{\b}}{$\mathcal{B}_\b$}
}
\idle{\t0}{\t0+\cc}{\ySa}{idle}
\foreach \b in {0,...,\NBmax}{%
  \pgfmathsetmacro{\xa}{\t0+\cc+\b*\ss}
  \tblock{\xa}{\xa+\ss}{\ySa}{\bfill{\b}}{$\mathcal{B}_\b$}
}
\pgfmathsetmacro{\xdrain}{\t0+\cc+\NBmax*\ss+0.5*\ss}
\node[ann, anchor=north, text=black!65] at (\xdrain,{\ySa-\lh-0.04}) {drain};

% ==================================================================
% card 2: without overlap
% ==================================================================
\draw[card] (\cardL,-3.15) rectangle (\cardR,-4.55);
\node[grouplab] at ({\cardL+\titleX},-3.21) {without overlap};
\node[lanelab] at ({\t0-0.06},\yCb) {Alltoallv};
\node[lanelab] at ({\t0-0.06},\ySb) {local SVDs};

\tblock{\t0}{\nbA}{\yCb}{black!12}{$\T[\widehat]{A}$}
\idle{\t0}{\nbA}{\ySb}{idle}
\tblock{\nbA}{\bEnd}{\ySb}{black!12}{$\T[\widehat]{A}$}

% ==================================================================
% what the overlap buys, and the time axis
% ==================================================================
\draw[guide] (\tEnd,-2.94) -- (\tEnd,-4.83);
\draw[guide] (\bEnd,-4.55) -- (\bEnd,-4.83);
\draw[meas] (\tEnd,-4.76) -- (\bEnd,-4.76);
\node[ann, anchor=north] at ({0.5*(\tEnd+\bEnd)},-4.79) {saved};

\draw[-{Stealth[length=1.4mm,width=1.2mm]}, draw=black!65, line width=0.45pt]
  (\t0,-5.13) -- (8.23,-5.13);
\node[ann, anchor=south west] at (\t0,-5.10) {time};

\end{tikzpicture}
  \caption{
  	Overlapping the redistribution of $\T[\widehat]{A}$ with slicewise SVDs, as seen by one processor.
  	}
  \label{fig:svd_overlap}
\end{figure}

\Cref{fig:svd_overlap} illustrates the benefit of the overlap.
The $\left(\prod_{k\ge 3} d_k\right)/p$ frontal slices the
processor will own are cut into $n_b$ batches, and each batch is exchanged
by its own non-blocking \texttt{MPI\_Ialltoallv} into a double-buffered
receive workspace. 
Each processor will contribute $1/p$th of $p \cdot \left(\prod_{k \ge 3} d_k \right) / n_b$ frontal slices to the collective, and it will receive $\left(\prod_{k \ge 3} d_k \right) / n_b$ complete frontal slices.
The two schedules carry the same pair of lanes so
they can be read against each other, and the dashed boxes mark the spans in
which the processor has nothing to compute. With overlap, the exchange for
batch $\mathcal{B}_b$ is posted and then flies while the slices of
$\mathcal{B}_{b-1}$, which have already arrived, are factored; the process
therefore stalls only for the first exchange, since nothing has arrived to
compute on yet, and the last batch drains with no communication left to hide
behind. Without overlap the entire exchange completes before any
factorization starts, so the stall lasts as long as the exchange does. The
schedule is illustrated as compute bound, the regime in which the overlap pays off;
when communication dominates, the later exchange blocks
extend past the SVDs below them and the cost of the SVD computations is hidden.

%\GB{The Ialltoallv has to happen over $n_b \cdot p$ columns from the matrix view, right?  The figure is showing only what the processor is receiving, not what it's sending.  I don't think we should change the figure to show this, I'm just not sure exactly what sentence to add to the text.  Maybe something that includes: ``all-to-alls are performed on data of size $(d_1d_2/p) \cdot n_b \cdot p=d_1d_2n_b$, which involves $n_b\cdot p$ frontal slices''?}

\subsection{Parallelizing Truncation for \tsvdmii}
\label{sec:truncation}

After each processor finishes the SVDs of its local frontal slices, it must determine how many singular values to keep. For \tsvdmi, this step is trivial: the user supplies a single rank $k$ and every slice is truncated to that rank. No communication is needed.

For \tsvdmii, the situation is different. The algorithm must choose a truncation rank $\rho_{f_i}$ for every frontal slice independently, subject to a user-specified relative error tolerance $\varepsilon$. This requires finding a global threshold singular value $\tau$ such that discarding every singular value below $\tau$ keeps the discarded spectral energy within the allowed budget. 
Because the singular values are distributed across processors, computing $\tau$ requires communication.

The simplest approach is to gather every singular value to a single root process, sort them, scan the sorted list to find $\tau$, and then broadcast $\tau$ back to every process. 
This is attractive because the number of singular values is small compared to the size of the tensor, and it works reasonably well in shared memory parallelization \cite{hussain2026highperformancestarmsvdbig}.
%: for an $N$-way tensor of shape $d_1 \times \cdots \times d_N$, there are only $\prod_{k=3}^N d_k$ singular values, one per frontal slice. For most practical tensors this is orders of magnitude smaller than the total number of entries, so the gather itself is cheap in absolute terms.

\begin{figure*}[t]
  \centering
  \input{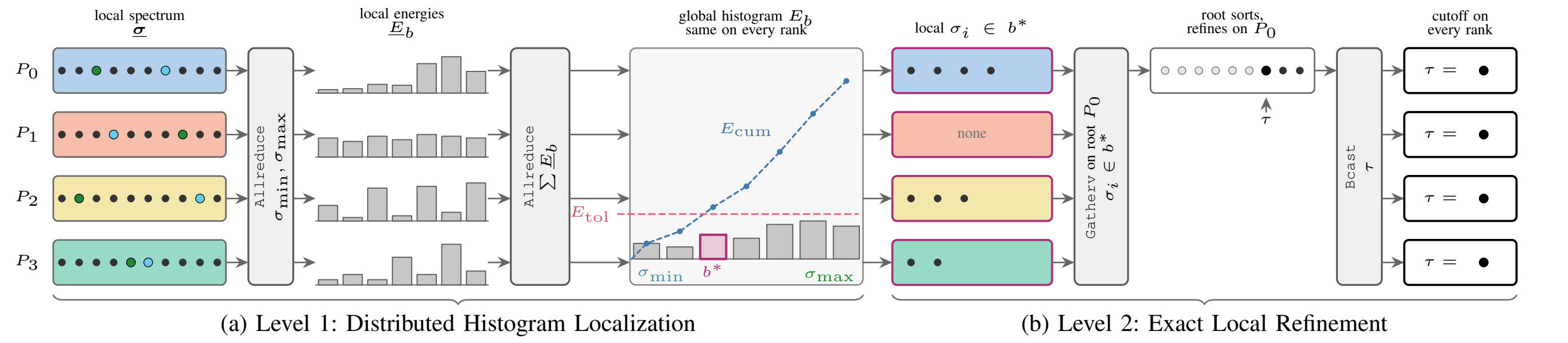}
  \caption{Two-level histogram refined cutoff on four processors (rows). 
  (a) Level 1: each processor accumulates spectral energy into $B$ logarithmically 
  spaced bins; a single \texttt{MPI\_Allreduce} produces a global histogram 
  requiring only $O(B)$ words per process. The first bin $b^*$ whose cumulative 
  energy exceeds the truncation budget $E_{\mathrm{tol}} = \varepsilon^2 \sum_i \sigma_i^2$ 
  (red dashed line) is identified as the target bin containing $\tau$. 
  (b) Level 2: only singular values falling inside $b^*$ are gathered to the root 
  via \texttt{MPI\_Gatherv}, sorted, and used to pinpoint the exact cutoff $\tau$, 
  which is then broadcast to all processors.}
  \label{fig:2lhrc}
\end{figure*}

However, as the tensor dimensions and node counts grow, the gather and sort on the root processor become a bottleneck. 
The root must store and sort every singular value in the entire tensor, and the time spent there can dominate the truncation phase. 
%Our experiments show that this full gather approach frequently takes seconds to over ten seconds, depending on the dataset.
To avoid gathering every singular value, we use a distributed two-level histogram refined strategy to find $\tau$. 
The idea is to compute local histograms, then communicate to obtain global spectral energies of each bin and determine in which bin the threshold $\tau$ lies.
Then only the singular values in the target bin need to be gathered and sorted for exact refinement.
\Cref{fig:2lhrc} illustrates the complete procedure on four processors.

\subsubsection{Level 1: Distributed Histogram Localization}
After selecting the parameter $B$, the number of histogram bins, the goal is to select splitters so that the number of global singular values in each bin is even.
To this end, we determine a global maximum and minimum via \texttt{MPI\_Allreduce}, and then we logarithmically space the bins between those values.
%Every processor computes its local minimum and maximum singular values, and an \texttt{MPI\_Allreduce} finds the global $\sigma_{\min}$ and $\sigma_{\max}$. The interval $[\sigma_{\min}, \sigma_{\max}]$ is partitioned into $B$ logarithmically spaced bins. 
Logarithmic spacing is important: the singular value spectra of compressible scientific tensors decay rapidly, so most of the energy lives in a small range near $\sigma_{\max}$, and uniform bins would waste resolution there. 
Each processor accumulates spectral energy $E_b = \sum_{\sigma_i \in b} \sigma_i^2$ into its local bins, and an \texttt{MPI\_Allreduce} produces a global histogram. 
The bins are scanned in ascending order to find the first bin $b^*$ whose cumulative energy exceeds the truncation budget. 
At this point, $\tau$ is known to lie inside $b^*$. This level is shown in \Cref{fig:2lhrc}(a).

\subsubsection{Level 2: Exact Local Refinement}
Only the singular values that fall inside $b^*$ are gathered to the root via \texttt{MPI\_Gatherv}. The root sorts this small candidate set and increments the cumulative energy until the budget is exceeded, selecting the corresponding singular value as the exact cutoff $\tau$. This value is then broadcast to every process, as shown in \Cref{fig:2lhrc}(b). Because only a small fraction of the singular values are communicated in the second level, the total communication volume is far smaller (by an expected factor of $O(B)$) than a full gather. 
Our experiments show that this approach reduces the truncation time by two to three orders of magnitude over the full gather.

Once $\tau$ is known, the physical truncation itself is entirely local. Each processor independently extracts the required columns from its local $\widehat{\mathbf{U}}$ and $\widehat{\mathbf{V}}$ factor matrices, keeping only those singular triplets whose singular value is at least $\tau$. 

\subsection{Full Parallel \tsvdmi and \tsvdmii Algorithms}
\label{sec:dist-tsvdm-algo}
%\GB{Add one pseudcode that explains the overlap and fixed-rank vs tolerance branching where tolerance branching would have the histogram thing.}
%\MTH{Srini, can you please own this subsection?}
%\MTH{
%	\begin{itemize}
%		\item We want to use MPI-ialltoall instead of reinventing new name.
%		\item Follow Naman's pseudocode closey to write the overlap section.
%		\end{itemize}
%}
\Cref{alg:partsvdm} is the pseudocode for both the parallel t-SVDM-I and t-SVDM-II algorithms described in the preceding sections (\cref{sec:partitioning,sec:alg:overlap,sec:truncation}). A separate listing (\cref{alg:histcutoff}) is provided for the parallel truncation phase for the t-SVDM-II algorithm. 

\begin{algorithm}[hbtp]
	\caption{Compute singular value threshold $\tau$}\label{alg:histcutoff}
	\begin{algorithmic}[1]
	\REQUIRE $\Pi$ is a $p \times 1$ grid of processors.
	\REQUIRE $\T[\widehat]{S} \in \Rn{d_1\times \cdots \times d_N}$ is a f-diagonal tensor distributed by linearizing the first two modes and evenly dividing them across the $p$ processors.
	\REQUIRE Energy tolerance $\epsilon$.
	\ENSURE $\tau \in \R$ is the threshold achieving $\epsilon$ relative error.
	\STATE {\textsc{HISTCUTOFF}}$(\Pi, \T[\widehat]{S}, \epsilon)$
	\STATE \COMMENT{Compute bin spacing}
	\STATE $\Vl{\sigma} \gets \textsc{concatenate}\pr{\mathrm{diag}\pr{\M[\widehat]{S}_{f_i}}} \forall f_i \in F$
	\STATE $\underline{\sigma}_{\min} \gets \min\pr{\Vl{\sigma}}$
	\STATE $\sigma_{\min} \gets \textsc{MPI\_Allreduce}\pr{\underline{\sigma}_{\min}, \text{MIN}}$ 
	\STATE $\underline{\sigma}_{\max} \gets \max\pr{\Vl{\sigma}}$	
	\STATE $\sigma_{\max} \gets \textsc{MPI\_Allreduce}\pr{\underline{\sigma}_{\max}, \text{MAX}}$
	\STATE $B \gets \textsc{logspace}(\sigma_{\min}, \sigma_{\max})$
	\STATE \COMMENT{Compute local histograms}
	\STATE $\underline{E}_b \gets \sum_{\sigma_i \in b} \sigma_i^2$
	\STATE $E \gets \textsc{MPI\_Allreduce}\pr{\underline{E}, \text{SUM}}$
	\STATE Find the first index $b^*$ such that $E_b >  \epsilon^2\sum_i \sigma_i^2$
	\STATE \COMMENT{Gather target bin to root}
	\STATE $\Vl{\sigma}_{b^*} \gets \textsc{concatenate}\pr{\sigma_i \in E_{b^*}}$
	\STATE $\V{s} \gets$\textsc{MPI\_Gatherv}(root, $\Vl{\sigma}_{b^*}$)
	\STATE \COMMENT{Compute exact local refinement and broadcast}
	\IF{root}
	\STATE Sort and scale $\V{s}$ based on global $E_{b^*}$
	\STATE Find $\tau$ such that $\V{s}_{\tau} \le \epsilon^2\sum_i \sigma_i^2$
	\STATE MPI\_Bcast($\tau$)
	\ENDIF
	\RETURN $\tau$
	\end{algorithmic}
\end{algorithm}

\begin{algorithm}[hbtp]
	\caption{Parallel $N$-Way t-SVDM Decomposition}\label{alg:partsvdm}
	\begin{algorithmic}[1]
		\REQUIRE $\Pi$ is a $p \times 1$ grid of processors. %\TODO{Srini, include both buffers for async sense. Add a line on packing block columns isntead of b1. Add mpi test to the pseudocode }
		\REQUIRE $\T{A} \in \Rn{d_1\times \cdots \times d_N}$ is distributed by linearizing the first two modes and evenly dividing them across the $p$ processors (see \cref{fig:distribution}(d)). $\Tl{A} \in \Rn{(d_1\cdot d_2) / p \times \cdots \times d_N}$ is the locally-owned tensor.
		\REQUIRE Transformation matrices $\Curly{\M{M}_i \in \Rn{d_i \times d_i}}_{i=3}^N$ replicated on all processors.
		\REQUIRE $r$ could either be an slicewise rank $k$ for t-SVDM-I or an energy tolerance $\epsilon$ for t-SVDM-II.
		\ENSURE $(\T[\widehat]{U}_k, \T[\widehat]{S}_k, \T[\widehat]{V}_k)$ is the t-SVDM computed whose frontal slices are stored evenly across the $p$ processors (see \cref{fig:distribution}(e)). $(\Tl[\widehat]{U}_k, \Tl[\widehat]{S}_k, \Tl[\widehat]{V}_k)$ are the locally-owned slices.
		\STATE {\textsc{par-t-SVDM}}$(\Pi, \T{A}, \Curly{\M{M}_i}, r)$
		\STATE \COMMENT{Local TTM to transform data}
		\STATE $ \Tl[\widehat]{A} \gets \Tl{A} \times_3 \M{M}_3 \times_4  \cdots \times_N \M{M}_N $
		\STATE \COMMENT{Redistribute frontal slices and factorize}
		\STATE prev $\gets$ none	
		\STATE rec $\gets$ none
		\STATE \COMMENT{Loop over all $n_b$ batches of column/slice blocks}
		\FOR{$i = 1$ to $n_b$}
			\STATE \COMMENT{Pack $i^\text{t}$ column/slice block}
			\STATE send $\gets$ \textsc{pack}($\Mz[\widehat]{\underline{A}}{1:2}[i]$)
			\STATE \COMMENT{Redistribute in non-blocking fashion}
			\STATE rec $\gets$ \textsc{MPI\_Ialltoallv}(send)
			\STATE \COMMENT{Overlap prev computation with communication}
			\IF{prev}
				\STATE \COMMENT{Process all slices in previous batch}
				\FOR{$f \in$ prev}
				\STATE $ \Ml[\widehat]{U}_f, \Ml[\widehat]{S}_f, \Ml[\widehat]{V}_f \gets \textsc{svd}(\Ml[\widehat]{A}_f)$ 
				\ENDFOR
			\ENDIF
			\STATE \textsc{MPI\_Wait}(rec)
			\STATE prev $\gets$ rec
		\ENDFOR
		%\PARFOR {$f \in f_B \text{ in batches of}~n_b$} 
		%\STATE $\textsc{ASYNCSEND}(f_B)$
		%\ENDPARFOR
		%\PARFOR {$(f \in f_B)$}
		%\STATE $\Tl[\widehat]{A} \gets \textsc{ASYNCRECV}(f_B)$
		%\STATE $ \Ml[\widehat]{U}_f, \Ml[\widehat]{S}_f, \Ml[\widehat]{V}_f \gets \textsc{svd}(\Ml[\widehat]{A}_f) $
		%\STATE $ \Tl[\widehat]{U}_f, \Tl[\widehat]{S}_f, \Tl[\widehat]{V}_f \gets \Ml[\widehat]{U}_f, \Ml[\widehat]{S}_f, \Ml[\widehat]{V}_f $
		%\ENDPARFOR
		\STATE \COMMENT{Compute slicewise ranks}
		\IF{$r$ is an integer}
			\STATE \COMMENT{All slices get $r$}
			\STATE $k_f \gets r$
		\ELSE
			\STATE \COMMENT{Compute global threshold via histogram refinement}
			\STATE $\tau \gets \textsc{HISTCUTOFF}(\Pi, \T[\widehat]{S}, r)$
			\STATE $\V{\rho}_f \gets 
			\#\left\{\M[\widehat]{S}_f > \tau \right\}$
			\STATE $k_f \gets \rho_f$
		\ENDIF
		\STATE \COMMENT{Truncate local factors}
		\FOR {$f \in \underline{F}$}
		%\STATE $ \Ml[\widehat]{U}_f, \Ml[\widehat]{S}_f, \Ml[\widehat]{V}_f \gets \textsc{truncate}(\Ml[\widehat]{U}_f, \Ml[\widehat]{S}_f, \Ml[\widehat]{V}_f, k_f) $
		\STATE $ \Tl[\widehat]{U}_k^f, \Tl[\widehat]{S}_k^f, \Tl[\widehat]{V}_k^f \gets \textsc{truncate}(\Ml[\widehat]{U}_f, \Ml[\widehat]{S}_f, \Ml[\widehat]{V}_f, k_f) $
		\ENDFOR
		% No need to retransform
		%\STATE $\Tl{U}_k \gets \Tl[\widehat]{U}_k \times_3 \mathbf{M}_3^{-1} \times_4 \cdots \times_N \mathbf{M}_N^{-1}$
		%\STATE $\Tl{S}_k \gets \Tl[\widehat]{S}_k \times_3 \mathbf{M}_3^{-1} \times_4 \cdots \times_N \mathbf{M}_N^{-1}$
		%\STATE $\Tl{V}_k \gets \Tl[\widehat]{V}_k \times_3 \mathbf{M}_3^{-1} \times_4 \cdots \times_N \mathbf{M}_N^{-1}$
		\RETURN $(\T[\widehat]{U}_k, \T[\widehat]{S}_k, \T[\widehat]{V}_k)$
	\end{algorithmic}
	\label{partsvdm}
\end{algorithm}

% !TEX root = ../main.tex

\section{Results}\label{sec:results}

\subsection{Experimental Setup}\label{sec:experimental_setup}

\distarm is implemented in C++ using MPI for inter-process communication along with BLAS and LAPACK for local computation.
For BLAS routines, we use Intel oneMKL~\cite{MKL} because it provides a batched BLAS interface, which is not part of standard BLAS libraries.
We use the CBLAS routines \texttt{?gemm} and \texttt{?gemm\_batch\_strided} for the TTMs and the LAPACK routines \texttt{?gesvd}, \texttt{?gesdd}, and \texttt{?gesvdx} for the slicewise SVDs.

We perform all experiments on the \texttt{highmemory} partition of the PARAM Pravega supercomputer~\cite{parampravega} at the Indian Institute of Science, using process-based parallelism only.
This partition has 156 nodes, each with two Intel Xeon Platinum 8268 processors (48 cores in total) and 768\,GB of memory.
The nodes are connected by HDR InfiniBand at 100\,Gb/s in a 1:1 fat-tree topology and share a 4\,PB Lustre filesystem.
We build our implementation with the Intel oneAPI \texttt{icpx} 2023.1.0 compiler, using the C++17 standard, the \texttt{-O3 -xHost} flags, and the C++ standard library of GCC 12.2.0.
We link against Intel MPI 2021.9.0 and the sequential LP64 interface of MKL 2023.2.0.
To time each phase, we place explicit barriers around it and take the maximum time over all processes.
For each configuration, we perform one warm-up run followed by five or ten timed runs and report the mean of these maxima.

%We run 48 MPI ranks per node and pin one rank to each physical core (\texttt{I\_MPI\_PIN\_PROCESSOR\_LIST=0-47}) with hyper-threading off, so each rank runs a single thread and calls sequential MKL.
%To limit variability between runs, we allocate nodes exclusively, exclude three nodes (\texttt{hm[113,138,139]}), and pin ranks as described above.
%Each phase is preceded by an \texttt{MPI\_Barrier} and timed as the maximum across ranks, and we report the mean of these maxima over the timed runs.

\subsection{Benchmark Datasets}\label{sec:datasets}
We evaluate our implementation on the five scientific datasets listed in \Cref{tab:benchmark_datasets}, which are 3- to 6-dimensional tensors ranging in size from 78\,GB to 768\,GB.

Three datasets come from fluid simulations.
The first, \dataset{miranda}, is the density field from a direct numerical simulation of the Rayleigh--Taylor instability, in which two fluids of relative density 1 and 3 undergo turbulent mixing~\cite{cabot2006reynolds}.
We use the full-resolution $3072^3$ version distributed by SDRBench~\cite{Zhao_2020,SDRBench}.
The second, \dataset{iso}, comes from the Johns Hopkins Turbulence Databases~\cite{yeung2012dissipation,li2008jhtdb,perlman2007jhtdb,jhtdb_isotropic4096}.
It holds one snapshot of forced isotropic turbulence at $Re_\lambda \approx 610$ on a $4096^3$ grid, and its three velocity components form a fourth mode.
The third, \dataset{halfcylinder}, is an ensemble of incompressible flows past a half-cylinder at Reynolds numbers 160, 320, 640, and 6400, simulated with the Gerris solver and resampled onto a regular grid~\cite{BaezaRojo19SciVisa,gerrisflowsolver}, with 151 time steps and three velocity components.

We also test our implementation on climate reanalysis and hyperspectral imaging data.
The climate dataset, \dataset{ncep-air}, is global air temperature from the NCEP/NCAR Reanalysis~\cite{kalnay1996ncep}, on a $2.5^\circ$ grid at 17 pressure levels.
\pystarm used a 10-year subset~\cite{hussain2026highperformancestarmsvdbig}; here we use 70 years (1948--2017), with time again split into time of day, day of year, and year, and we remove February~29 so that every year has 365 days.
The imaging dataset, \dataset{icvl}, contains 186 hyperspectral images of natural scenes~\cite{arad_and_ben_shahar_2016_ECCV}, each with 31 spectral bands from 400 to 700\,nm .

To isolate the effect of individual design decisions, we additionally use six synthetic 4-way tensors of 64\,GB each (\texttt{float32}), whose frontal slices range from $64 \times 64$ to $2048 \times 2048$: $64 \times 64 \times 2048 \times 2048$, $128 \times 128 \times 1024 \times 1024$, $256 \times 256 \times 512 \times 512$, $512 \times 512 \times 256 \times 256$, $1024 \times 1024 \times 128 \times 128$, and $2048 \times 2048 \times 64 \times 64$.
They are used in the overlap study (\Cref{sec:overlap}) and the truncation study (\Cref{tab:truncation_comparison}).
Their entries are drawn independently and uniformly from $[-100, 100]$.

\begin{table}[t]
\caption{Scientific datasets used for evaluating the distributed-memory \tsvdmi and \tsvdmii implementations.\label{tab:benchmark_datasets}}
\centering
\footnotesize
\setlength{\tabcolsep}{4pt}
\renewcommand{\arraystretch}{1.2}
\begin{tabular}{%
  l
  l
  l}
\hline
\textbf{Dataset} &
\textbf{Dimensions} &
\textbf{Size} \\
\hline

\dataset{miranda} &
$3072 \times 3072 \times 3072$ &
108 GB (\texttt{float32}) \\

\dataset{halfcylinder} &
$640 \times 240 \times 80 \times 151 \times 3 \times 4$ &
83 GB (\texttt{float32}) \\

\dataset{icvl} &
$1392 \times 1300 \times 31 \times 186$ &
78 GB (\texttt{float64}) \\

\dataset{ncep-air} &
$73 \times 144 \times 17 \times 4 \times 365 \times 70$ &
136 GB (\texttt{float64}) \\

\dataset{iso} &
$4096 \times 4096 \times 4096 \times 3$ &
768 GB (\texttt{float32}) \\

\hline
\end{tabular}
\end{table}

\subsection{Comparison with Existing Shared-Memory Solution}\label{sec:pystarm_comparison}
\begin{figure}[t]
    \centering
    \includegraphics[width=\columnwidth]{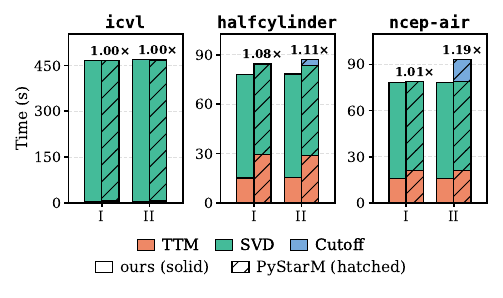}
    \caption{Single-node comparison against \pystarm, the existing
    shared-memory implementation, on one PARAM Pravega node (48 cores). Solid
    bars are \distarm running 48 MPI ranks on that node; hatched
    bars are \pystarm using 48 threads. Both implementations use \texttt{float64}. I is \tsvdmi
    at full rank and II is \tsvdmii with $\varepsilon = 10^{-2}$. Bars are
    broken down by phase, and the number above each \pystarm\ bar gives its
    time relative to ours for the same dataset and algorithm. 
    %All times exclude MPI-I/O and the final truncation of the factors. 
    The \dataset{miranda} and
    \dataset{iso} datasets are absent because \pystarm\ runs out of memory on them.}
    \label{fig:pystarm_comparison}
\end{figure}

We first compare against \pystarm~\cite{hussain2026highperformancestarmsvdbig}, the existing shared-memory implementation of \tsvdmi and \tsvdmii.
\pystarm runs on a single node, so we run both implementations on one node.
\pystarm supports only \texttt{float64}, so both implementations use \texttt{float64} here.
\Cref{fig:pystarm_comparison} shows \tsvdmi at full rank and \tsvdmii with $\varepsilon = 10^{-2}$ on \dataset{icvl}, \dataset{halfcylinder}, and \dataset{ncep-air}.

On a single node, the two implementations have comparable runtime.
The slicewise SVD phase dominates in both and takes a similar time.
In \distarm, this phase also includes the overlapped all-to-all redistribution.
The differences come from the TTM and cutoff phases.
Our TTM phase is faster, by up to $1.9\times$ on \dataset{halfcylinder}.
\pystarm applies each TTM to one large buffer that holds the whole tensor.
In \distarm, each MPI rank applies it to its own smaller local buffer.
While both implementations use MKL's batched strided GEMM for the TTM, we suspect that the larger buffer leads to poorer use of the cache.
The \tsvdmii cutoff takes 0.8 to 14.4\,s in \pystarm and at most 0.11\,s in ours.
This is because \pystarm computes the cutoff serially, while \distarm parallelizes the computation.
We evaluate our parallel cutoff in more detail in \cref{sec:truncation_results}.

A main limitation of the shared-memory implementation is memory, as all data must fit on one node.
\pystarm ran out of memory on \dataset{miranda} and \dataset{iso}.
%For \dataset{ncep-air}, we use the entire 70 years of data.
\pystarm can fit all 70 years on one node, but its internal memory use is high (about $4\times$ the 136\,GB input), so it failed to run on our platform.
\distarm is not limited to the memory of one node and scales well to 32 nodes on this dataset, as we show in \cref{sec:strong_scaling}.

\subsection{Overlapping Communication with SVD Computation}\label{sec:overlap}
\begin{figure}[t]
    \centering
    \includegraphics[width=\columnwidth]{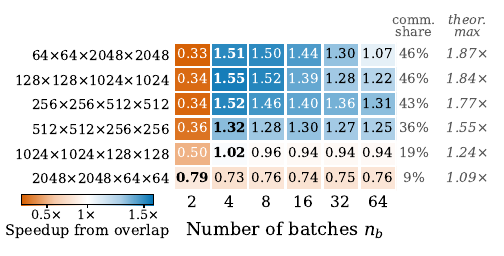}
    \caption{Speedup from overlapping the redistribution with the slicewise
    SVDs, defined as the runtime without overlap divided by the runtime with
    overlap, as the number of batches $n_b$ varies, on 32 nodes (1536 MPI
    ranks). All six tensors are 64\,GB; their frontal slices range from $64 \times 64$ to
    $2048 \times 2048$, and rows are ordered by slice size, smallest first.
    The first column on the right gives the share of the non-overlapped
    runtime spent in the all-to-all communication, which falls from
    communication-heavy at the top to compute-bound at the bottom; the second,
    in italics, gives the theoretical maximum speedup that perfect overlap
    could reach, the non-overlapped runtime divided by the larger of its
    communication and SVD times. Blue cells mark a gain, orange cells
    a loss, and the best $n_b$ for each tensor is in bold.}
    \label{fig:overlap_speedup}
\end{figure}

\Cref{fig:overlap_speedup} shows the effect of overlapping the redistribution with the slicewise SVDs, as described in \cref{sec:alg:overlap}, on the six synthetic tensors, run on 32 nodes (1536 MPI ranks, 48 per node).
When communication takes nearly half of the total runtime, as in the top rows of \cref{fig:overlap_speedup}, we save time by hiding the communication behind the SVD computations.
When communication has a smaller share, as in the bottom rows, the overhead of overlapping increases the runtime.

Given the relative amount of time spent communicating without overlap, we calculate the maximum speedup that perfect overlap could reach.
The last column of \cref{fig:overlap_speedup} reports this maximum, so it can be compared with the observed speedup.
Note that at most $2\times$ speedup is possible with overlapping, which happens when communication has $50\%$ share of the runtime.
In this experiment, the data size (and the amount of communication) is fixed while the size of the frontal slices is varied.
When the size of the frontal slices is small, the cost of the SVDs is lower and the relative cost of communication is higher.
Larger frontal slices lead to a smaller share of communication.

\subsection{Performance of \tsvdmii Truncation Strategy}\label{sec:truncation_results}
We compare two approaches to find the global cutoff $\tau$ for \tsvdmii (\cref{sec:truncation}).
The full gather (FG) baseline gathers all singular values to a root process, which sorts them to find $\tau$, similar to the approach of \pystarm.
Our two-level histogram refined cutoff (2LH) first locates the bin containing $\tau$ using a global histogram, and then gathers and sorts only the singular values in that bin (\cref{fig:2lhrc}).

\Cref{tab:truncation_comparison} shows that 2LH is faster than FG by more than two orders of magnitude on all tensors.
In FG, all singular values are sent to a single root process in one large \texttt{MPI\_Gatherv}, and the root then sorts all of them while the other processes wait.
In 2LH, all processes bin their own singular values in parallel, and a single \texttt{MPI\_Allreduce} communicates only one value per bin.
Then only the singular values in the target bin $b^*$ are gathered to the root and sorted (\cref{fig:2lhrc}).
Thus 2LH reduces both the communication volume and the sorting volume.
If $B$ bins are used with perfect balance across bins, the cost of communicating and sorting is reduced by approximately a factor of $B$.
On the synthetic tensors considered in \cref{tab:truncation_comparison}, the number of singular values halves from one row to the next, and the times of both approaches roughly halve with it.

\begin{table}[t]
\caption{Cutoff localization time for the Full Gather (FG) baseline against
the two-level histogram refined cutoff (2LH) with $B = 2048$ histogram bins,
on 32 nodes (1536 MPI ranks) with $\varepsilon = 10^{-2}$ and the DCT transform. Note the differing units:
FG is reported in seconds, 2LH in milliseconds.\label{tab:truncation_comparison}}
\centering
\footnotesize
\setlength{\tabcolsep}{4pt}
\renewcommand{\arraystretch}{1.2}
\begin{tabular}{l r r r}
\hline
\textbf{Tensor} & \textbf{FG (s)} & \textbf{2LH (ms)} & \textbf{Speedup} \\
\hline
$64 \times 64 \times 2048 \times 2048$ & 30.678 & \textbf{19.35} & $1585\times$ \\
$128 \times 128 \times 1024 \times 1024$ & 15.174 & \textbf{9.53} & $1592\times$ \\
$256 \times 256 \times 512 \times 512$ & 7.115 & \textbf{5.86} & $1214\times$ \\
$512 \times 512 \times 256 \times 256$ & 3.389 & \textbf{2.87} & $1181\times$ \\
$1024 \times 1024 \times 128 \times 128$ & 1.644 & \textbf{1.47} & $1118\times$ \\
$2048 \times 2048 \times 64 \times 64$ & 0.745 & \textbf{0.89} & $840\times$ \\
\hline
\dataset{miranda} & 1.164 & \textbf{0.76} & $1532\times$ \\
\dataset{halfcylinder} & 3.816 & \textbf{1.56} & $2446\times$ \\
\dataset{icvl} & 0.991 & \textbf{2.73} & $363\times$ \\
\dataset{ncep-air} & 14.849 & \textbf{4.57} & $3250\times$ \\
\hline
\end{tabular}
\end{table}

\subsection{Strong Scaling of Distributed \tsvdmi and \tsvdmii}\label{sec:strong_scaling}

\begin{figure*}[t]
    \centering
    \includegraphics[width=\textwidth]{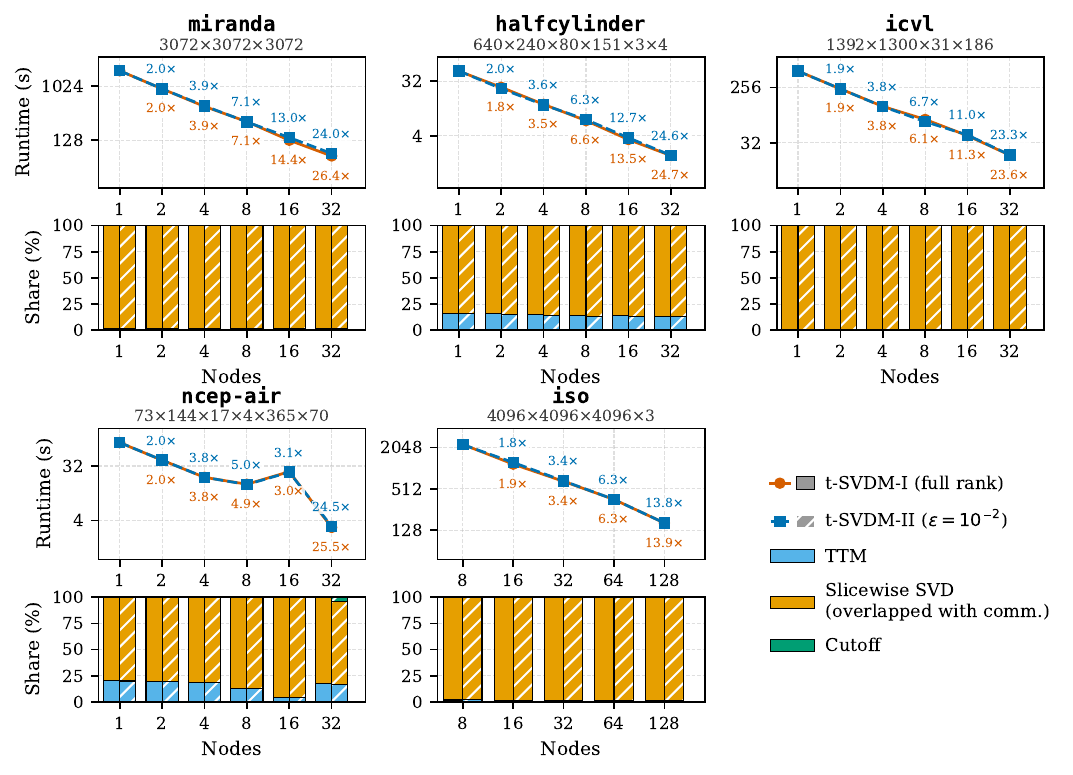}
    \caption{Strong scaling of \tsvdmi (full rank) and \tsvdmii ($\varepsilon = 10^{-2}$) on \dataset{miranda}, \dataset{halfcylinder}, \dataset{icvl}, and \dataset{ncep-air} from 1 to 32 nodes, and on \dataset{iso}, which does not fit on fewer than 8 nodes, from 8 to 128 nodes. Top: total runtime, annotated with the speedup over the smallest node count (\tsvdmi below the line, \tsvdmii above). Bottom: share of runtime spent in the TTM and slicewise SVD phases for \tsvdmi (solid) and \tsvdmii (hatched); the slicewise SVD phase includes the all-to-all communication overlapped with it. The \tsvdmii cutoff phase takes under 0.2\% of runtime and is not visible at this scale.}
    \label{fig:strong_scaling_combined}
\end{figure*}

\Cref{fig:strong_scaling_combined} shows that both \tsvdmi and \tsvdmii scale well on all datasets.
Every dataset keeps at least 70\% parallel efficiency at the largest node count.
This is because the initial TTM decouples the tensor into independent frontal slices.
The compute-intensive SVDs on these slices then run in parallel, which scales very well.
The redistribution involves all of the data across all the processors, but we are able to hide most of that cost behind the SVDs.
Note that \dataset{iso} does not fit on fewer than 8 nodes, so we measure its speedup relative to 8 nodes.

The slicewise SVD phase dominates the runtime on all datasets.
This phase also includes the all-to-all communication, which we overlap with the SVD computation (\cref{sec:overlap}).
On \dataset{ncep-air} with 32 nodes, overlapping makes the full run $1.55\times$ faster for \tsvdmi and $1.43\times$ faster for \tsvdmii.

The TTM phase takes a visible share of the runtime only on \dataset{halfcylinder} and \dataset{ncep-air}.
These 6-way tensors need TTMs along middle modes, which can be less efficient than TTM along the first or last mode due to the access pattern.
Still, the TTM share does not grow with the number of nodes, so it does not limit scalability.

% \begin{itemize}
% 	\item As we are dominated by SVD, overlap optimization is key to get good scaling. We can say something  like - when communication has large share as communication is not expected to scale as we go to larger scale.
% 	\MTH{Naman, can you add the numbers for real dataset for overlap and non-overlap cases?}
% \NP{Climate data (\dataset{ncep-air}) on 32 nodes, times in seconds with overlap off / on:\\
% \parbox{\linewidth}{\vspace{3pt}\centering\footnotesize\setlength{\tabcolsep}{3pt}\begin{tabular}{l ccc ccc}
% \hline
%  & \multicolumn{3}{c}{SVD} & \multicolumn{3}{c}{Total} \\
%  & off & on & speedup & off & on & speedup \\
% \hline
% \tsvdmi  & 4.17 & 2.52 & $1.65\times$ & 4.75 & 3.06 & $1.55\times$ \\
% \tsvdmii & 3.90 & 2.53 & $1.54\times$ & 4.57 & 3.19 & $1.43\times$ \\
% \hline
% \end{tabular}\vspace{3pt}}}
% 	\item {More tensor dimension means ttm on the middle mode so bandwidth is coming into play.}
% 	\item {For all cases we are getting 70\% efficientcy.}
% 	\item {Iso data did not fit in one node so we started from 8 nodes.}
% 	\end{itemize}

\subsection{Compression Performance}\label{sec:compression_results}

% \begin{itemize}
% 	\item DCT transformation compresses better.
% 	\item $10^-3$ gives 10x compression.
% 	\item $10^-1$ is reasonable error for visualization and it gives orders of magnitude compression.
% 	\item tsvdm-ii gives space to compress more.
% 	\item tsvdm-ii compression is comparable to \tuckermpi and put a forward reference to the Tucker comparison section.
% \end{itemize}

\begin{figure}[t]
    \centering
    \includegraphics[width=0.95\columnwidth]{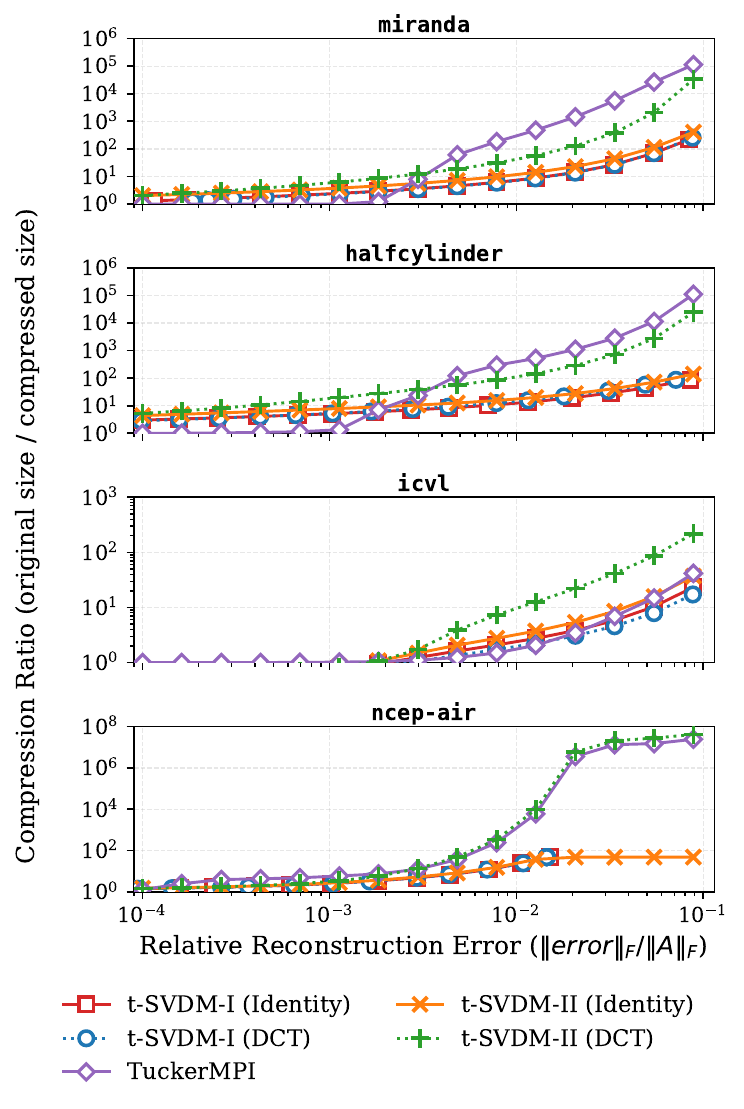}
    \caption{Compression ratio versus relative reconstruction error for the \tsvdmi and \tsvdmii algorithms using both Identity and DCT transforms across the scientific datasets, with \tuckermpi for comparison. Each panel has its own y-axis range.}
    \label{fig:compression_combined}
\end{figure}

\Cref{fig:compression_combined} shows the compression ratio against the relative reconstruction error for all datasets.

We first observe that the DCT compresses better than the identity transform across all 4 datasets, especially with \tsvdmii.
The DCT captures the correlation along the transformed modes, which opens up more room for compression.
The identity transform does not capture this correlation.

\tsvdmii also compresses better than \tsvdmi, as expected, because \tsvdmi keeps the same rank in every frontal slice.
Since the rank of \tsvdmi cannot go below one, the compression of \tsvdmi levels off once every slice reaches rank one--\dataset{ncep-air} shows this most clearly.
\tsvdmii chooses the rank of each slice from a single global cutoff, so it can drop low-energy slices entirely.
This is also why \tsvdmii gains the most from the DCT at looser tolerances.

\begin{figure*}[t]
    \centering
    \includegraphics[width=\textwidth]{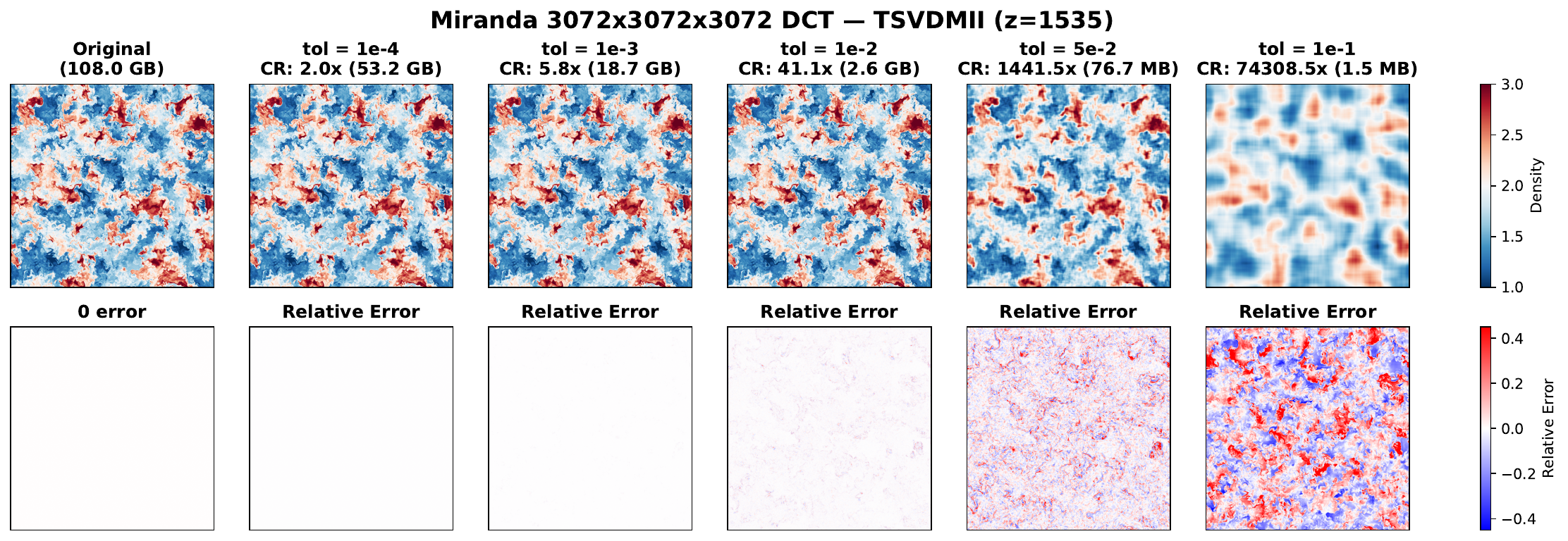}
    \caption{Reconstruction of slice 1535 of \dataset{miranda} by \tsvdmii with the DCT at different error tolerances. The top row shows the original and reconstructed slices, and the bottom row shows the relative error. At $10^{-2}$, fine-scale features are preserved with $41.1\times$ compression. At $10^{-1}$, the large-scale structures remain at only 1.5 MB.}
    \label{fig:reconstruction_miranda_tsvdmii}
\end{figure*}

Choosing the right error tolerance depends on the downstream task involving the compressed data.
For visualization only, we expect an error in the range of $10^{-1}$ to be sufficient.
At this error, \tsvdmii compresses by orders of magnitude, beyond $10^4\times$ on all datasets except \dataset{icvl}.
The reconstruction of the \dataset{miranda} dataset shown on the right side of \cref{fig:reconstruction_miranda_tsvdmii}, for example, still keeps the large-scale structures, although fine details are lost.

\begin{figure}[t]
    \centering
    \includegraphics[width=\columnwidth]{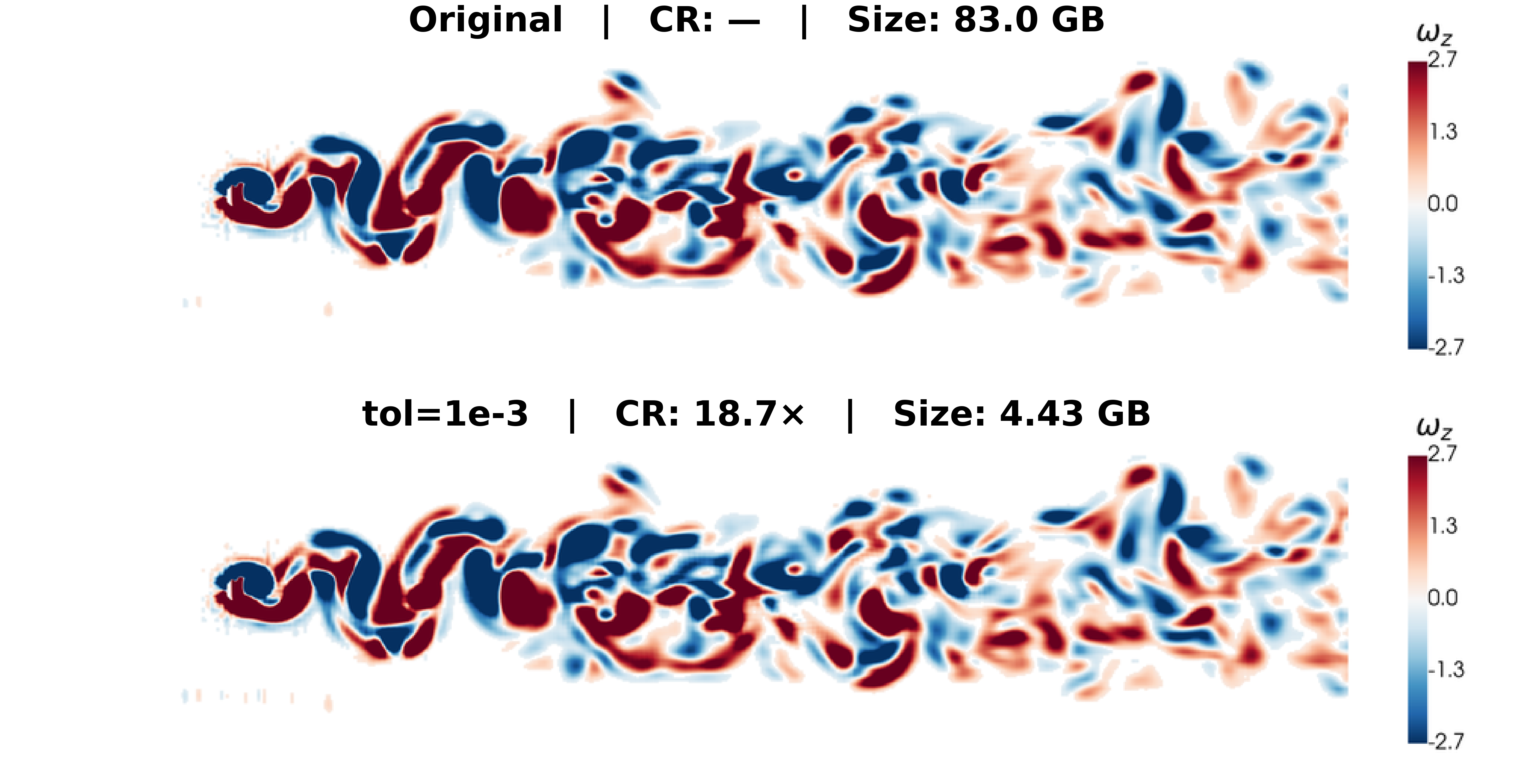}
    \caption{The $z$-component of the vorticity field of \dataset{halfcylinder} ($Re=6400$, time step 150), computed from the velocity. Top: original data. Bottom: \tsvdmii reconstruction with the DCT at an error of $10^{-3}$. The two are visually indistinguishable.}
    \label{fig:reconstruction_vorticity}
\end{figure}

We expect tighter error tolerances to be required for scientific computation tasks beyond visualization.
\Cref{fig:reconstruction_miranda_tsvdmii} shows the difference in visualization of a slice of the \dataset{miranda} tensor across the various tolerances.
At error of $10^{-3}$, \tsvdmii with the DCT compresses most datasets by about $10\times$.
The only exception is \dataset{icvl}, which shows less compressibility overall.
The reconstructions at this error are visually indistinguishable from the original data and can be used for downstream computational tasks.
\Cref{fig:reconstruction_vorticity} shows a harder case.
The \dataset{halfcylinder} tensor stores only the velocity.
We compute the vorticity $\omega_z = \partial v/\partial x - \partial u/\partial y$ by differentiating the reconstructed velocity field.
Derivatives tend to be sensitive to the errors; thus vorticity is a stricter test of the efficacy of compression.
At $10^{-3}$ error, the vorticity from the reconstruction is visually indistinguishable from the original, suggesting that the compression can potentially support downstream tasks that rely on derivatives of the underlying field.

%\MTH{How is the marker so regularly spaced?}
%\NP{I used 20 point between 1e-4 and 1e-0, and took them using numpy.logspace.}
%\GB{But what about TSVDM-I?  Those markers shouldn't fall in perfect vertical alignment with TSVDM-II's should they?}

%Finally, the compression of \tsvdmii is comparable to that of \tuckermpi, which we discuss in \cref{sec:tuckermpi_comparison}.

\subsection{Comparison with \tuckermpi}\label{sec:tuckermpi_comparison}

\tuckermpi~\cite{ballard2020tuckermpi} is a popular distributed memory parallel implementation used to compute a Tucker tensor decomposition, which is also used for data compression.
We compare \tsvdmii with \tuckermpi in terms of compression, performance, and scalability on four of our scientific datasets.
\tuckermpi requires the user to choose an $N$-way processor grid, which has a large effect on its runtime.
We chose grids using manual tuning starting from suggested heuristics for \tuckermpi~\cite[Section~9.4.2]{ballard2020tuckermpi} and list them in \cref{tab:tuckermpi_grids}.

\begin{figure}[t]
    \centering
    \includegraphics[width=\columnwidth]{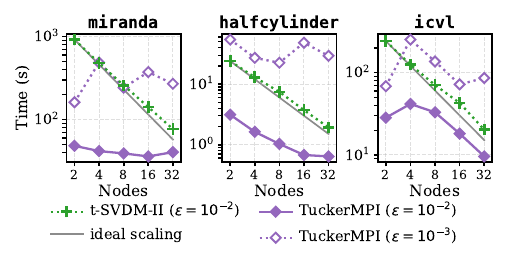}
    \caption{Strong scaling of \tsvdmii (DCT, $\varepsilon=10^{-2}$) and \tuckermpi ($\varepsilon=10^{-2}$ and $10^{-3}$) on \dataset{miranda}, \dataset{halfcylinder}, and \dataset{icvl}: computation time (excluding I/O) against the number of nodes, both on log scales. The grey line is ideal scaling from the 2-node \tsvdmii time.}
    \label{fig:tucker_vs_tsvdmii_scaling}
\end{figure}

\Cref{fig:compression_combined} shows that the compression behavior of Tucker and Star-M SVD approximations are comparable.
Neither approach is always superior to the other across all datasets or relative errors, and the choice of matrix $\textbf{M}$ can have a large effect on compression ratio for Star-M SVD.
We observe in \cref{fig:tucker_vs_tsvdmii_scaling} that the performance of \tsvdmii is also comparable to \tuckermpi.
For \tuckermpi, the computation and communication, and therefore the runtime, vary a lot with the tolerance, because it uses an algorithm with the feature that higher compression leads to lower computation time.
In contrast, the runtime of our \tsvdmii implementation depends less on the tolerance, because it computes the full SVD of each slice regardless of the tolerance.
We also note that \tuckermpi requires specifying a processor grid configuration that has a large effect on running time; we empirically tuned these parameters and report the best performance observed.
For a loose error tolerance (high compression), \tuckermpi is faster than \tsvdmii, and for tight error tolerance (low compression), \tsvdmii is faster.
We observe good scalability for \tsvdmii for all three datasets, while the scalability for \tuckermpi varied over datasets and tolerances.

\begin{table}[t]
\caption{Processor grids used for the \tuckermpi strong-scaling runs in \cref{fig:tucker_vs_tsvdmii_scaling}, with 48 MPI ranks per node.\label{tab:tuckermpi_grids}}
\centering
\footnotesize
\setlength{\tabcolsep}{4pt}
\renewcommand{\arraystretch}{1.2}
\begin{tabular}{r c c c}
\hline
\textbf{Nodes} & \dataset{miranda} & \dataset{halfcylinder} & \dataset{icvl} \\
\hline
2  & $1 \times 4 \times 24$  & $1 \times 2 \times 4 \times 12 \times 1 \times 1$  & $1 \times 1 \times 8 \times 12$ \\
4  & $1 \times 8 \times 24$  & $1 \times 2 \times 8 \times 12 \times 1 \times 1$  & $1 \times 1 \times 12 \times 16$ \\
8  & $1 \times 16 \times 24$ & $1 \times 4 \times 8 \times 12 \times 1 \times 1$  & $1 \times 2 \times 12 \times 16$ \\
16 & $1 \times 24 \times 32$ & $1 \times 4 \times 16 \times 12 \times 1 \times 1$ & $1 \times 4 \times 12 \times 16$ \\
32 & $1 \times 32 \times 48$ & $1 \times 8 \times 16 \times 12 \times 1 \times 1$ & $1 \times 4 \times 12 \times 32$ \\
\hline
\end{tabular}
\end{table}

\section{Conclusions}\label{sec:conclusions}

This paper presented the first distributed-memory MPI implementation of the \tsvdmi and \tsvdmii tensor decomposition frameworks within the $\star_M$ algebraic setting. To address the unique computational and communication challenges of these algorithms, we introduced several key optimizations: a zero-copy batched tensor-times-matrix (TTM) formulation that eliminates data movement, a double-buffered overlapped execution strategy that hides the latency of global tensor redistribution behind facewise SVD computations, and a scalable two-level log-histogram cutoff strategy that reduces the adaptive truncation phase to milliseconds.

Extensive evaluations on large-scale scientific datasets (up to 768~GB) demonstrated robust strong scaling, achieving parallel efficiencies of 73\%--83\% on 32 nodes (1536 MPI tasks), and of 86\%--87\% on 128 nodes (6144 MPI tasks), relative to 8 nodes, for the 768~GB \dataset{iso} dataset. In terms of compression performance, \tsvdmii with the DCT transform consistently outperformed \tsvdmi. Our framework utilizes a fixed one-dimensional data distribution that requires no parameter tuning while maintaining high parallel efficiency.

By bridging the gap between the theoretical optimality of transform-domain tensor decompositions and the practical demands of high-performance computing, this work establishes distributed \tsvdmi and \tsvdmii as scalable, robust tools for the compression and analysis of massive scientific datasets.

\balance

\section*{Acknowledgments}

This material is based upon work supported by the U.S. Department of Energy, Office of Science, Office of Advanced Scientific Computing Research, under Awards DE-SC0023296 and DE-SC0025394 (G. Ballard and M.T. Hussain) and under Contract DE-AC02-06CH11357 at Argonne National Laboratory (S. Eswar and V. Rao).
N. Pesricha acknowledges Master's Research Fellowship from the Ministry of Education, Government of India.
P. Motamarri acknowledges financial support from the Google India Research Award and ARG MATRICS (Grant No.\ ANRF/ARGM/2025/002824/TS) from the Anusandhan National Research Foundation during the course of this work.
We acknowledge the National Supercomputing Mission (NSM) for providing the computing resources of PARAM Pravega at the Supercomputer Education and Research Centre (SERC), Indian Institute of Science.
We thank the Johns Hopkins Turbulence Databases for the isotropic turbulence data, and NOAA PSL, Boulder, Colorado, USA, for the NCEP/NCAR Reanalysis data.
We used Anthropic's Claude Opus 5.5 to help draw the figures and improve the prose of this manuscript.
The authors take full responsibility for every statement in this paper, whether or not it was generated with AI assistance.

% {\appendix[Proof of the Zonklar Equations]
% Use $\backslash${\tt{appendix}} if you have a single appendix:
% Do not use $\backslash${\tt{section}} anymore after $\backslash${\tt{appendix}}, only $\backslash${\tt{section*}}.
% If you have multiple appendixes use $\backslash${\tt{appendices}} then use $\backslash${\tt{section}} to start each appendix.
% You must declare a $\backslash${\tt{section}} before using any $\backslash${\tt{subsection}} or using $\backslash${\tt{label}} ($\backslash${\tt{appendices}} by itself
%  starts a section numbered zero.)}

% {\appendices
% \section*{Proof of the First Zonklar Equation}
% \section*{Proof of the Second Zonklar Equation}
% Appendix two text goes here.}

% \section{References Section}
% You can use a bibliography generated by BibTeX as a .bbl file.
%  BibTeX documentation can be easily obtained at:
%  http://mirror.ctan.org/biblio/bibtex/contrib/doc/
%  The IEEEtran BibTeX style support page is:
%  http://www.michaelshell.org/tex/ieeetran/bibtex/
 
% \bibliography{IEEEabrv,../bib/paper}
% You can manually copy in the resultant .bbl file and set second argument of $\backslash${\tt{begin}} to the number of references
%  (used to reserve space for the reference number labels box).

\pagebreak
\bibliographystyle{IEEEtran}
\bibliography{references}

% \newpage


\vfill

\end{document}